\newcommand{\CLASSINPUTtoptextmargin}{1.75cm}
\newcommand{\CLASSINPUTbottomtextmargin}{1.75cm}
\newcommand{\CLASSINPUToutersidemargin}{1.7cm}

\documentclass[10pt,twocolumn,journal]{IEEEtran}

\usepackage{amsmath, mathrsfs, mathtools}
\usepackage{amsfonts}
\usepackage{epstopdf}
\usepackage{amssymb}
\usepackage{amsthm}
\usepackage{dsfont}
\usepackage{graphicx}
\usepackage{color}
\usepackage{graphicx}
\usepackage{subfig}
\usepackage{caption}
\usepackage[backend=bibtex,giveninits=true,sorting=none]{biblatex}
\usepackage{balance}
\renewcommand*{\bibfont}{\footnotesize}

\let\labelindent\relax
\usepackage{enumitem}

\newcommand{\subparagraph}{}
\usepackage[compact]{titlesec}
\titlespacing*{\section}{15pt}{1.2\baselineskip}{0.9\baselineskip}

\theoremstyle{remark}

\newtheorem{remark}{\bf Remark}

\long\def\comment#1{}

\newcommand\figref{Figure~\ref}

\newcommand{\ben}{\begin{enumerate}}
\newcommand{\een}{\end{enumerate}}

\newcommand{\beq}{\begin{equation}}
\newcommand{\eeq}{\end{equation}}

\newcommand{\bi}{\begin{itemize}}
\newcommand{\ei}{\end{itemize}}

\DeclareMathOperator*{\argmax}{arg\,max}

\newcommand{\CC}{\mathbb{C}}
\newcommand{\PP}{\mathbb{P}}
\newcommand{\RR}{\mathbb{R}}

\newcommand{\EE}{\mathbb{E}}

\newcommand{\av}{{\bf a}}

\newcommand{\cv}{{\bf c}}

\newcommand{\hv}{{\bf h}}

\newcommand{\rv}{{\bf r}}
\newcommand{\sv}{{\bf s}}

\newcommand{\xv}{{\bf x}}
\newcommand{\yv}{{\bf y}}
\newcommand{\zv}{{\bf z}}

\newcommand{\Am}{{\bf A}}
\newcommand{\Bm}{{\bf B}}
\newcommand{\Cm}{{\bf C}}

\newcommand{\Hm}{{\bf H}}
\newcommand{\Id}{{\bf I}}

\newcommand{\Rm}{{\bf R}}
\newcommand{\Sm}{{\bf S}}

\newcommand{\Xm}{{\bf X}}
\newcommand{\Ym}{{\bf Y}}
\newcommand{\Zm}{{\bf Z}}

\newcommand{\Cc}{{\cal C}}

\newcommand{\Kc}{{\cal K}}
\newcommand{\Lc}{{\cal L}}

\newcommand{\tauv}{\hbox{\boldmath$\tau$}}

\newcommand{\Gammam}{\boldsymbol{\Gamma}}

\newcommand{\SINR}{{\sf SINR}}
\newcommand{\SNR}{{\sf SNR}}

\newcommand{\herm}{{\sf H}}

\title{Pilot-Based Unsourced Random Access with a Massive MIMO Receiver: Interference Cancellation and Power Control}
\author{Alexander Fengler, Osman Musa, Peter Jung, Giuseppe Caire
\thanks{Parts of this paper have been accepted for presentation at the
    IEEE International Workshop on Signal Processing Advances in
Wireless Communications (SPAWC) 2021 \cite{Fen2021e}.}
\thanks{The authors are with the Communications and Information Theory Group,
Technische Universit\"{a}t Berlin (\{fengler, osman.musa, peter.jung,
caire\}@tu-berlin.de).}
}

\begin{document}

\maketitle

\begin{abstract}
    In this work we treat the unsourced random access problem on a Rayleigh
    block-fading AWGN channel with multiple receive antennas. Specifically, we
    consider the slowly fading scenario where the coherence block-length is
    large compared to the number of active users and the message can be
    transmitted in one coherence block. Unsourced random access refers to a
    form of grant-free random access where users are considered to be a-priori
    indistinguishable and the receiver recovers a list of transmitted messages
    up to permutation.  In this work we show that, when the coherence block
    length is large enough, a conventional approach based on the transmission
    of non-orthogonal pilot sequences with subsequent channel estimation and
    Maximum-Ratio-Combining (MRC) provides a simple energy-efficient solution
    whose performance can be well approximated in closed form. Furthermore, we
    analyse the MRC step when successive interference cancellation (SIC) 
    is done in groups, which allows to strike a balance between receiver complexity
    and reduced transmit powers. Finally, we investigate the impact of power control
    policies taking into account the unique nature of massive random access, including
    short message lengths, uncoordinated transmission, a very large amount of concurrent 
    transmitters with unknown identities, channel estimation errors and decoding errors.
    As a byproduct we also present an extension of the MMV-AMP algorithm which allows to treat
    pathloss coefficients as deterministic unknowns by performing maximum likelihood estimation
    in each step of the MMV-AMP algorithm.
\end{abstract}


\section{Introduction}
Conventional random access protocols in current mobile communication standards establish
an uplink connection between user and base station (BS) by first running a multi-stage handshake protocol \cite{Dah2013,Dah2018}.
During this initial access phase active users are identified and subsequently
a scheduler assigns orthogonal transmission resources to the active users.
One of the paradigms of modern machine-type communications \cite{Tal2012} consists of a very large
number of devices (here referred to as ``users'') with sporadic data. Typical examples thereof are
Internet-of-Things (IoT) applications, wireless sensors deployed to monitor smart infrastructure,
and wearable biomedical devices \cite{Has2013}. In such scenarios, a BS should be
able to collect data from a large number of devices. However, due to the sporadic nature of the
data generation and communication, the initial access procedure is overly wasteful.

An alternative way of communication is known as grant-free random access, where users transmit their
data without awaiting the grant of transmission resources by the BS. A commonly discussed 
grant-free strategy in a massive multi-user MIMO setting is to assign 
fixed orthogonal or non-orthogonal pilot sequences to users
\cite{Lar2014,Mar2016,Liu2018a,Liu2018b,Che2018,Liu2018c,Sen2018a}. Active users then transmit their pilot sequence
directly followed by their data sequence.
The BS identifies the active users in the first step and estimates their channel vectors. Subsequently
the channel estimates are used to detect the data sequences using either maximum-ratio-combining (MRC)
or zero-forcing \cite{Mar2016}.

However, as the number of users in a system grows large and the access frequencies become small, it gets
increasingly inefficient to assign fixed pilot sequences to all the users.
In contrast, {\em unsourced random access} (U-RA) is a novel grant-free paradigm proposed in \cite{Pol2017}
and motivated by an IoT scenario where millions of cheap devices have their codebook hardwired at
the moment of production, and are then disseminated into the environment.  In this case,  all users
make use of the very same codebook and the BS decodes the list of transmitted messages
irrespectively of the identity of the active users.
The U-RA approach can simplify the random access protocol design because it does not require an initial
access phase and, in contrast to existing grant-free approaches, it allows for a system that is completely independent of the inactive users, which makes
it well suited to the IoT scenario with a huge number of devices with very sporadic activity. 

Practical coding schemes for U-RA have been mainly studied on the real AWGN channel,
e.g. \cite{Pol2017,Ord2017,Fen2021d,Cal2020,Vem2017,Kow2019c,Ama2020a,Pra2020a,Ust2019},
and on the Rayleigh quasi-static fading AWGN channel \cite{Kow2020,And2020a}.  
The U-RA problem on a Rayleigh block-fading AWGN channel in a massive MU-MIMO setting was formulated
in \cite{Fen2021a} and it was shown that a covariance-based activity detection (AD) algorithm combined with a tree code \cite{Vem2017}
can achieve sum-spectral-efficiencies that grow proportional to the coherence block-length $n$,
even if the number of active users is significantly larger than $n$,
specifically up to $K_a = \mathcal{O}(n^2)$.

In typical wireless systems the coherence block-length $n$ may range from
a couple of hundred to a couple
of thousand, depending mainly on the speed of the transmitters. At a carrier frequency of
$2$ GHz the coherence times, according to the model $T_c \approx 1/(4 D_s)$ where $D_s$ is the maximal Doppler
spread \cite{Tse2005}, may range from 
$45$ ms at $3$ km/h to $1$ ms at $120$ km/h. The coherence bandwidth depends on the maximal delay spread and,
in an outdoor environment,
typically ranges from
$100$ to $500$ kHz depending on the propagation conditions.
Therefore, the number of complex symbols in an OFDM coherence block
may range from $n=100$ to $n=20000$, depending mainly on the assumed speed and the geometry of the environment.
Unfortunately, the covariance-based AD algorithm in \cite{Fen2021a} has a run-time complexity
that scales with $n^2$, which makes it unfeasible to use at $n \gtrsim 300$. So while the covariance based approach
of \cite{Fen2021a} is well suited to a fast-fading scenario with $n\lesssim 300$, it becomes unfeasible for
large coherence block-lengths in the order of thousands. On the other hand,
algorithms based on hybrid-GAMP \cite{Shy2020a} and tensor-based modulation \cite{Dec2020c}
have shown an excellent performance
at large coherence block-length at which the covariance-based approach is no longer feasible. 

In this work we present a conceptually simple algorithm that can be used when $n>K_a$. It is based
on pilot transmission, AD, channel estimation, MRC and single-user decoding, very similar
to the state-of-the art approach for massive MIMO grant-free random access \cite{Liu2018a,Liu2018b,Sen2018a}.
In contrast to the scheme with fixed pilots allocated to all users, we use a pool of non-orthogonal pilots
from which active users pick one pseudo-randomly based on the first bits of their message.
We show that a collision of users, i.e. two users picking the same pilot sequence,
can be resolved by using a polar single-user code with a successive-cancellation-list (SCL) decoder.  
Finite-length simulations shows that the performance of the coding scheme can be well
predicted by analytical
calculations. Despite its simplicity the suggested scheme has an energy efficiency that is comparable to
existing approaches. 
Note, that the problem treated here is formally
almost equivalent to grant-free random access with fixed pilots allocated to each user. Differences arise only
in the possibility of collisions and the associated use of an list decodable single-user code. 
The error probability of AD and MRC in the asymptotic limit $K_a,K_\text{tot},n \to \infty$ with fixed ratios $K_a/K_\text{tot}$ and
$K_\text{tot}/n$ has been analysed in \cite{Liu2018b}.
In this work we focus on the finite-blocklength regime
and the combination of MRC with a single-user polar code. Besides, we investigate simplified
successive interference cancellation (SIC) receivers which allow to relax the power control
requirements and/or decrease the average transmit power while still maintaining a manageable 
decoding complexity. The capacity of grouped SIC was studied in \cite{Han1995} and it was
found that when all users are received with the same power it is possible to get arbitrary close to the Shannon capacity
with $\mathcal{O}(1)$ cancellation steps by adapting the rates of the users. In this work we develop
a framework that allows to calculate the number of groups required to achieve a given per-user
error probability taking into account channel estimation errors and decoding errors.

Previous works on URA assumed either a perfect constant receive-power control
policy, i.e. that all received signal strengths (RSS) are equal
\cite{Kow2020,And2020a,Dec2020c,Fen2021a} or no power control, i.e. the RSSs
follow some distribution that has a random shadowing component and a distance
dependent pathloss component \cite{Shy2020a}. In current cellular networks
power control is handled either in an open loop fashion, where users aim to
equalize their received power based on e.g. downlink (DL) pathloss estimates, or in
a closed control loop, where the BS sends power adjustment commands to each
user until each user has a satisfactory SINR.  Clearly, closed loop power
control is infeasible in an mMTC system because individually managing the
powers of several hundred users would lead to an unacceptable delay compared to
the short message length. On the other hand, open loop power control is often
imperfect and the pathloss estimation error at the devices has to be considered in the
modelling. A no-power-control policy does have some advantages in an mMTC
scenario.
First, there
is no delay associated with a power control phase, second, the transmit antenna
does not need to have a large dynamic range, which can lead to cheaper devices, and third, 
the imbalance in received powers can be used as an additional source of diversity to separate users,
a concept known as power-domain non-orthogonal multiple-access (NOMA) \cite{Hig2015,Isl2017}.
However, the downside is that
the overall power consumption may increase, because strong users will spend more power than necessary and, in addition,
cause more interference for other users.
In this work we introduce a partial power control scheme that can minimize the average transmit power,
even below the
value which would be needed with an equal receive-power policy, while still
allowing for simplified transmitter design.
The partial power control scheme relaxes perfect power control by introducing a
discrete set of received power levels.  Active users then dynamically choose
their transmit power to target one of those power levels based on their large-scale fading coefficient (LSFC),
assuming that they know their LSFC from DL measurements.  In general it is hard
to optimally choose the received power levels since the amount of users at a
certain level is random and users choose their power level independent of each
other. Nonetheless, in an mMTC scenario where the number of active users is
large, e.g. several hundred, the fraction of users per level will sharply
concentrate around their average values, provided that the number of levels is
not too large.  This concentration, which is unique to a massive random access setting allows
for the design of optimized power control policies.

\subsection{Contribution and related works}
\begin{itemize}
    \item We introduce a unsourced random access coding scheme for the quasi-static Rayleigh fading channel.
    \item We evaluate a simplified grouped SIC strategy at the receiver which allows to active users to transmit
        without power control while still maintaining a low enough receiver complexity. 
    \item We introduced ML estimation of LSFCs within MMV-AMP which allows to estimate the received power levels
        in a non-Bayesian way. The method is significantly faster than Bayesian posterior-mean-estimation while achieving
        a comparable activity detection error.
    \item We give an analysis of the effective SINRs at the receiver taking into account SIC, imperfect channel estimation,
        short blocklengths
        and the random nature
        of uncoordinated random access.
    \item We introduce a partial power control policy where users are received at one of $G$
        possible power levels and use the analysis to optimise the power levels.
\end{itemize}

Power control in MU-MIMO has been studied e.g. in \cite{Mar2016,Sen2018a,Ngo2015b}, also in the context of
NOMA \cite{Hig2015,Isl2017}.
The problem at hand has some unique properties.
For once, the rates of all users are fixed in the URA model, therefore no rate adaption is possible. The massive amount
of users precludes any use of feedback based power control policies, even in its simplest form,
e.g. 1-bit UP/DOWN power commands.
In \cite{Yat1995} Yates gave an iterative method for optimizing the received powers for
a large class of standard interference functions. In a MU-MIMO system the channel estimation error
depends inversely on the transmit power in the pilot phase.
Therefore an increase in power of one user may decrease
the interference for some of the users, which breaks the monotonicity assumption of Yates' standard
interference function.
\footnote{We assume here that the transmit power of a user in the pilot and data phase is the same.
In principle the pilot powers should be optimized as well which is out of the scope of this paper.}
Power control in conjunction with
imperfect SIC has been studied in \cite{And2003,Agr2005} for CDMA.
In \cite{Agr2005} an algorithm was given that allows
to find the minimal received power levels such that all the SINRs are equal after SIC.
The drawbacks of the algorithm of \cite{Agr2005} is that the power of each user
is controlled individually
which is hardly achievable with uncoordinated transmission.
Furthermore, only the received power is minimized and not the
transmit power. 

MMV-AMP has been proposed in \cite{Kim2011} and its performance as 
an AD algorithm for MU-MIMO has been analysed in \cite{Liu2018a,Liu2018b,Liu2018c,Che2018}.
In \cite{Che2018} unknown LSFCs were handled in a Bayesian optimal way by including the distribution
of the LSFCs in the posterior mean estimator (PME). This approach has two downsides: for once, it is computationally
expensive when the PME cannot be calculated in closed form, and second, common models for the
distribution of LSFCs are mostly empirical and often unreliable. An alternative is to use a soft-thresholding function
as denoiser \cite{Liu2018c}.
The downside of the soft-thresholding function is that it depends on a parameter which has to be optimized
empirically. A similar approach has been used with the related hybrid-GAMP algorithm in \cite{Bel2018a},
where a Laplace prior distribution is assumed for the LSFCs. In the SMV case it is known that
a Laplace prior leads to a soft-thresholding denoising function and that AMP with a
soft-thresholding denoiser
is asymptotically equivalent to the popular LASSO estimator which is known to be min-max optimal \cite{Bay2012}.
In this work we present an new
approach that performs a ML estimation of the LSFCs in each MMV-AMP iteration. This approach treats
the LSFCs as deterministic unknowns and does not require any prior information on the distribution
of the LSFCs. We show that the proposed approach performs well in practice
without the need for optimizing parameters.
\section{Channel model}
\label{sec:model}
We consider a quasi-static Rayleigh fading channel with 
a block of $n$ signal dimensions over which the user channel vectors are constant.
In contrast to the block-fading channel treated in \cite{Fen2021a}, were a message is encoded over multiple independent
fading blocks, here we assume that a message can be transmitted
in a single coherence block. 
Following the problem formulation in \cite{Pol2017}, each user is given the same codebook
$\Cc = \{ \cv_m : m \in [2^{nR}]\}$, formed by 
$2^{nR}$ codewords $\cv_m \in \CC^n$. 
The codewords are normalized such that $\|\cv_m\|^2_2 \leq n$.
An unknown number $K_a$ out of $K_{\rm tot}$ total users transmit their
message over the coherence block.
Let $\mathcal{K}_a$ denote the set of active users, $i_k$ denote the index of the message chosen by user $k$,
$h_{k,m}\sim\mathcal{CN}(0,1)$ iid be the Rayleigh channel coefficient between user $k$ and receive antenna $m$
and let $g_k\in\RR_+$ denote the large-scale fading coefficient (LSFC) of user $k$, which captures the path-loss
and shadowing components of the fading. Let $P_k$ denote the power-per-symbol of user $k$. The received signal strength (RSS) is then given by $P_kg_k$.
Furthermore, let $\gamma_k = g_k$ for $k\in \mathcal{K}_a$ and
zero otherwise. The received signal at the $m$-th receive antenna takes the form
\beq
\yv_m =\sum_{k=1}^{K_\text{tot}}\sqrt{P_k\gamma_k}h_{k,m} \cv_{i_k} + \zv_m = \sum_{k\in\mathcal{K}_a} \sqrt{P_k g_k} h_{k,m} \cv_{i_k} + \zv_m 
\eeq 
where $z_{m,i}\sim\mathcal{CN}(0,N_0)$ iid.  
The BS must then produce a list $\Lc$ of the transmitted messages $\{m_k : k\in \Kc_a\}$ (i.e., 
the messages of the active users). Let $n_\text{md} = \sum_{k \in \mathcal{K}_a} \mathds{1}_{\{m_k \notin \mathcal{L}\}}$ denote the number
of transmitted messages missing in the output list.
The system performance is expressed in terms of the {\em Per-User Probability of Misdetection}, defined as
\beq
p_{md} = \frac{\EE\left[ n_\text{md}  \right]}{K_a}
\label{eq:ura_pmd}
\eeq
where the expectation is taken over the random choices of codewords, the fading and the noise.
In applications a slight overhead in the list size may be tolerable if it reduces the misdetections.
Since the number of active users is not necessarily known, it is practical to let the decoder decide
on a list size, which therefore becomes a random variable. 
Let $n_\text{fa} = |\mathcal{L}\setminus \{ m_k : k \in \Kc_a \}|$
denote the number of messages in the output list that were not transmitted by any user, also called
\emph{False Alarms}.
$n_\text{fa}$ is related to the list size by
\beq
|\mathcal{L}| = n_\text{fa} + K_a - n_\text{md}
\label{eq:fa_md_relation}
\eeq 
To get an empirical performance measure we define  
the {\em Probability of False-Alarm} as the average fraction of false alarms, i.e., 
\beq
p_{fa} = \EE\left[\frac{n_\text{fa}}{|\mathcal{L}|}\right].
\label{eq:ura_pfa}
\eeq
Operationally, $p_\text{fa}$ is the probability that a randomly chosen message from the output list
is a false alarm.
In the special case where $K_a$ is known at the receiver and $|\mathcal{L}| = K_a$ is fixed it follows from \eqref{eq:fa_md_relation}
that $p_\text{fa} = p_\text{md}$.
$p_\text{fa}$ and $p_\text{md}$ are also referred to as per-user probabilities of error (PUPEs).

Notice that in this problem formulation the number of total users $K_{\rm tot}$ is completely irrelevant,
as long as it is much larger than the range of possible active user set sizes $K_a$
(e.g., we may consider $K_{\rm tot} = \infty$). 
Furthermore, as
customary in coded systems, we express energy efficiency in terms of the
standard quantity $E_b/N_0 :=  \frac{P}{R N_0}$. 

In line with the classical massive MIMO setting \cite{Mar2016},
we assume an independent Rayleigh fading model for the channel coefficients $h_{k,m}$,
such that the channel vectors
for different users are independent from each other and are spatially white (i.e.,
uncorrelated along the antennas), that is,  $\hv_k = (h_{k,1},...,h_{h,m})^\top \sim \mathcal{CN}(0, \Id_M)$.

\subsection{Power Control}
\label{sec:power_control}
In a multiple access system without SIC it is usually desired that active users are received at the same power level
at the receiver.
This can be achieved, possibly up to some error, through a DL beacon signal which
all devices listen to to estimate their LSFC. Then they can adjust their transmit power accordingly.
It is well known though, e.g. in CDMA systems, that a non-uniform received power distribution is beneficial
when SIC is used at the receiver side \cite{Ver1998a}. Therefore, in this work we investigate the advantages of
non-uniform received power distributions in a massive random access system. Specifically in an mMTC
scenario a relaxation of perfect power control can lead to several desirable effects.
1) A reduced or simplified power control phase can reduce the access latency of the active devices
2) If SIC is used at the receiver, a heterogeneous received power distribution may lead to a reduced
required average transmit power
3) When the transmitter does not have to fully invert its LSFC its dynamic range can be reduced, which
allows for cheaper amplifiers at the transmitter side.

Throughout this work we assume the LSFCs of active users are distributed according to the shadowing-pathloss
model
\beq
g_k [\text{dB}] = -\alpha - \beta\log_{10}(d_k) + \sigma^2_\text{shadow}z,
\label{eq:shp}
\eeq
where $d_k$ is the distance
from user $k$ to the BS in km and $z \sim \mathcal{N}(0,1)$. The active user are then allowed to adapt
their transmit power based on their LSFC, which we assume they know from downlink measurements.
This leads to different distributions of received powers depending on the power control policy.
We investigate the following policies in this work
\begin{itemize}
    \item No power control (NPC): All users have the same transmit power. The distribution of received powers is given by a scaled
        version of \eqref{eq:shp}. In addition, we limit the maximal received power to $g_\text{max}$, which can be achieved
        in practice by limiting the minimal distance to the BS.
    \item Statistical channel inversion (SCI): User perfectly invert their LSFC. All users are received
        at the same power level $P_kg_k = \text{const.}$. 
    \item Imperfect SCI: As above, but some error is allowed, i.e. because of measurement errors.
    \item Partial SCI: Users invert their channel not fully but only to reach one of $G$ possible
        levels. Users choose this level based on their own LSFC independent of the other users
        $P_kg_k \in\{\pi_1,...,\pi_G\}$. The average fraction of users on level $i$ is denoted
        by $\xi_i$.
\end{itemize}

\section{Pilot-based massive MIMO U-RA}
Let the coherence block be divided into two periods of lengths $n_p$ and $n_d$.
In the first period each user chooses one of $N = 2^J$ (non-orthogonal) pilot sequences based on the first
$J$ bits of its message.
Let $\Am \in \CC^{n_p\times N}$ denote the matrix of pilot sequences with columns normalized as $\|\av_i\|_2^2 = n_p$
and for $i=1,...,N$ let $\mathcal{A}_i$ denote the set of users that have chosen the pilot with index $i$.
The received signal in the identification phase can be written in matrix form as
\begin{align}\label{pilot_sig}
    \Ym_p = \Am \tilde{\Gammam} \tilde{\Hm} + \Zm_p \in \CC^{n_p \times M},
\end{align}
$\tilde{\Hm} \in \CC^{N\times M}$ is a matrix with iid $\mathcal{CN}(0,1)$ entries
and $\tilde{\Gammam}$ is the diagonal matrix with
$\tilde{\gamma}_i = \sum_{k\in\mathcal{A}_i}P_{k}g_k$ on the diagonal. Note, that we have used the
Gaussianity of the channel vectors here,
which implies that $\sum_{k\in\mathcal{A}_i} \sqrt{P_kg_k}\hv_k \sim \mathcal{CN}(0,\tilde{\gamma}_i\Id_M)$.
We also introduce the indicator variables $\tilde{b}_i$ which indicate whether the pilot with index $i$ has been chosen
by at least one user: 
\beq
\tilde{\gamma}_i =: \tilde{b}_i\tilde{g}_i
\label{eq:tilde_g}
\eeq
where $\tilde{b}_i = 1$ if $\mathcal{A}_i \neq \emptyset$ and $0$ otherwise.
Note, that channel vectors are only defined for those indices which have been chosen by the active users.
Formally, we define the remaining channel vectors as zero.
The BS uses an AD algorithm as in \cite{Fen2021a} to estimate $\mathcal{I}$, the indices of
the used pilots, and the corresponding LSFCs.
Let $\hat{\Gammam}$ denote the matrix with the estimates $\hat{\gamma}_k$ of $\tilde{\gamma}_k$ on the diagonal and let 
$\hat{\mathcal{I}}$ be the estimate of the set of active pilots. 
For an index set $\mathcal{I}$ and for any matrix $\Bm$ let $\Bm_\mathcal{I}$
denote the matrix that contains only the columns of $\Bm$ with indices in $\mathcal{I}$. 
Then a linear MMSE estimate of the channel matrix is computed as
\beq
\hat{\Hm} = \hat{\Gammam}_{\hat{\mathcal{I}}}^{1/2}\Am_{\hat{\mathcal{I}}}^\herm \left(\Am_{\hat{\mathcal{I}}}\hat{\Gammam}_{\hat{\mathcal{I}}}\Am_{\hat{\mathcal{I}}}^\herm + N_0\Id_{n_p}\right)^{-1}\Ym_p \in \CC^{\hat{K}_a \times M}
\label{eq:lmmse}
\eeq 
where $\Am_{\hat{\mathcal{I}}}$ denotes a sub-matrix of the pilot matrix $\Am$ which contains only the columns
which have been estimated as active and $\hat{\Gammam}_{\hat{\mathcal{I}}}$ contains the
estimates of $\tilde{\gamma}_i$ on the diagonal. 
In the second period each users encodes its remaining $B-J$-bit message with a binary $(B-J,2n_d)$
block code and modulates
the $2n_d$ coded bits via QPSK on a sequence of $n_d$ complex symbols $\sv_k$.
These are transmitted over the $n_d$ channel uses
in the second phase. The matrix of received signals in the second phase is
\beq 
\Ym_d = \sum_{k\in\mathcal{K}_a} \sqrt{P_kg_k}\sv_k\hv_k + \Zm_d \in \CC^{n_d \times M}.
\eeq 
The BS uses the channel estimate $\hat{\Hm}$ from the first phase to perform user separation, or more precisely message separation,
via MRC,
i.e. it computes 
\beq
\hat{\Sm} = \hat{\Gammam}_{\hat{\mathcal{I}}}^{-1/2}\hat{\Hm}\Ym_d^\herm \in \CC^{\hat{K}_a \times n_d}
\eeq 
The rows of $\hat{\Sm}$ correspond to estimates of the transmitted sequences $\sv_k$.
Note, that it is also possible to use zero-forcing \cite{Mar2016}
instead of MRC but this would require that $M>K_a$.
The rows of $\hat{\Sm}$ are individually demodulated,
the bit-wise log-likelihood ratios are computed and fed into a soft-input single-user decoder. 
If the decoder finds a valid codeword, the index of the corresponding pilot is converted back to a
$J$ bit sequence and prepended to the codeword. Then the combination of the two is added to the output list.
The use of a polar code with CRC-bits and a successive-cancellation-list decoder has the additional benefit
that we can include all the valid codewords in the output list of the SCL decoder in the U-RA output list.
This allows to recover the messages of colliding users which have chosen the same pilot in the first phase.
The ability of polar codes to resolve sums of codewords has been observed and used for U-RA on the AWGN
in combination with spreading sequences \cite{Pra2020a} and a slotted Aloha approach \cite{Mar2019,And2020a}.
After a valid codeword has been recovered it can be re-encoded and subtracted from $\Ym_d$, followed
by a new MRC and single-user decoding attempt. This is known
as successive interference cancellation (SIC). Besides full SIC (each correctly
recovered codewords gets cancelled individually) and no SIC, we consider a grouped SIC approach were
groups of users are decoded in parallel before all correctly decoded codewords from this group
are cancelled. Note, that the CRC-bits of the polar code allow to detect with high probability
whether decoding was successful.
\subsection{Activity Detection with MMV-AMP}
\label{sec:mmv_amp}

For AD in the pilot phase we use the MMV-AMP algorithm, which was
introduced in \cite{Kim2011}, and used for 
AD in a Bayesian setting
where the LSFCs are either known, or their distribution is known \cite{Liu2018a,Che2018}. 
The algorithm aims to recover the unknown matrix
$\Xm = \tilde{\Gammam}\tilde{\Hm}$
from the linear Gaussian measurements $\Ym_p$ defined in \eqref{pilot_sig}. 
Let $\Xm_{k,:}$ denote the $k$-th row of $\Xm$.
The MMV-AMP 
iterations are defined as follows: 
\begin{align}
    \tau^2_{t+1,i} &= \frac{\|\Zm^t_{:,i}\|_2^2}{N} \quad i = 1,...,M\\
    \Xm^{t+1} &= \eta(\Am^\herm \Zm^t + \Xm^t,\tauv_{t+1})     \label{eq:vamp_1} \\
    \Zm^{t+1}   &= \Ym_p - \Am\Xm^{t+1} + \frac{N}{n_p} \Zm^{t}
    \langle \eta^\prime(\Am^\herm \Zm^{t} + \Xm^{t},\tauv_{t+1})\rangle       \label{eq:vamp_2} 
\end{align}
with $\Xm^0 = 0$ and $\Zm^0 = \Ym_p$. The function
$\eta: \CC^{N\times M}\times\RR^{M} \to \CC^{N\times M}$ is defined row-wise as
\begin{equation}
    \eta(\Rm,\tauv) =
    \left [ \begin{array}{c} \eta_{1}(\Rm_{1,:},\tauv) \\ \vdots \\ 
\eta_{{N}}(\Rm_{N,:},\tauv) \end{array} \right ],  \label{etamatrix}
\end{equation}
where each row function $\eta_{k}: \CC^M\times\RR^M \to \CC^M$ is chosen as 
an estimate of the random vector $\xv_k = \sqrt{\tilde{\gamma}_k}\hv_k$
in the {\em decoupled} Gaussian observation model
\beq
\rv_k = \xv_k + \zv_k,
    \label{eq:vamp_decoupled}
\eeq
where $\zv_k$ is a complex Gaussian noise vector with components $\sim \mathcal{CN}(0,\text{diag}(\tauv))$.
The choice of $\eta$ will be discussed later.
The term $\langle\eta^\prime(\cdot,\cdot)\rangle$ in \eqref{eq:vamp_2} is defined as
\beq
\langle\eta^\prime(\Rm,\tauv)\rangle 
= \frac{1}{N}\sum_{k=1}^{N} \eta_{k}^\prime(\Rm_{k,:},\tauv), 
\eeq
where $\eta_{k}^\prime(\cdot,\tauv) \in \CC^{M\times M}$
is the  Jacobi matrix of the function $\eta_{k}(\cdot,\tauv)$ evaluated at the $k$-th row $\Rm_{k,:}$ of the matrix argument $\Rm$.
Calculating the full matrix $\eta^\prime_{k}(\rv,\tauv)$ at each iteration would give a complexity per-row per-iteration
of $\mathcal{O}(M^2)$. 
Note, that for many commonly used $\eta$ the diagonal terms are much larger than the off-diagonal terms,
Therefore, we approximate the derivative by calculating only the diagonal elements of 
$\eta^\prime_{k}(\rv,\tauv)$ and setting the rest to zero, as described in \cite{Fen2021a}.

\subsection{The choice of $\eta$}
Two approaches for the choice of $\eta$ in the AD setting have been suggested in the literature \cite{Liu2018a,Che2018}.
In the Bayesian optimal version of MMV-AMP $\eta_k$ are chosen as the posterior mean estimates (PME) of
$\xv_k$ 
over the joint distribution of
$\xv_k$ and
$\zv_k$ defined by the decoupled channel \eqref{eq:vamp_decoupled} and the prior distribution of $\xv_k$.
\beq
    \eta_k(\rv_k,\tauv) = \EE[\xv_k|\rv_k]
    \label{eq:uncond_eta}
\eeq 
The other approach is to assume that $\tilde{g}_k$ are known from prior measurements and
calculate the conditional PME over the joint distribution of $\tilde{\hv}_k$ and $\zv_k$.
\begin{equation} \label{PME-cond}
    \eta_{k}(\rv,\tilde{g}_k,\tauv) := \EE[ \xv | \rv, \tilde{g}_k].
\end{equation}
It is important to recall the difference between the AD setting and the URA setting here. 
In the AD setting each column of $\Am$ is permanently associated with
a user, which in principle allows to measure the LSFCs of those users beforehand.
In contrast, in URA the columns of $\Am$ are associated with messages which are randomly chosen
by active users. Therefore, it is conceptually impossible to obtain 
the coefficients $\tilde{g}_i$ in \eqref{eq:tilde_g} beforehand, even if the LSFCs of all users are known.
In addition, the distribution of $\tilde{g}_i$ is very complicated due to the possibility of multiple users picking the same pilot,
which makes the Bayesian optimal MMV-AMP infeasible here. A valid approach to approximate the distribution of $\tilde{\gamma}_i$
when $K_a \ll N$ is to neglect the possibility of collisions, which leads to 
\begin{align}
    p(\tilde{\gamma}_i) &\approx p(|\mathcal{A}_i| = 0)\delta(\tilde{\gamma}_i) + p(|\mathcal{A}_i| = 1)p_g(\tilde{\gamma}_i = P_{\mathcal{A}_i}g_{\mathcal{A}_i}) \label{eq:su_approx}\\
                        &= \left(1-N^{-1}\right)^{K_a}\delta(\tilde{\gamma}_i) \\
                        &\quad +\left[1-\left(1-N^{-1}\right)^{K_a}\right]p_g(\tilde{\gamma}_i = P_{\mathcal{A}_i}g_{\mathcal{A}_i})\\
                    &= \left(1-\frac{K_a}{N}\right)\delta(\tilde{\gamma}_i) \\
                    &\quad +\frac{K_a}{N}p_g(\tilde{\gamma}_i = P_{\mathcal{A}_i}g_{\mathcal{A}_i}) + \mathcal{O}\left(\frac{K_a^2}{N^2}\right)
\end{align}
where $p_g(\cdot)$ denotes the distribution of the LSFCs and $\delta(x) = 1$ if $x = 0$ and $0$
otherwise. In \cite{Fen2021d} it was shown
that in the unfaded single-antenna case AMP with the mismatched PME using this
approximated prior converges rapidly to the Bayesian optimal AMP.  With the
approximation \eqref{eq:su_approx} the PME in the URA setting reduces to the PME in
the AD setting. Nonetheless, even with such an approximation using the PME in
MMV-AMP comes with some severe drawbacks: the distribution of LSFCs is often complicated
and varies over several orders of magnitude. In addition, for most commonly used distributions the PME can not be computed
in closed form, so it has to be computed by numerical integration which
significantly increases the computational complexity. In the following we suggest an alternative approach
that treats factors $\tilde{\gamma}_k$ as deterministic unknowns and does not require any
statistical knowledge of the LSFCs.
\subsection{Maximum Likelihood Estimation of the LSFCs}
\label{sec:lsfc_est}
We suggest to compute an ML estimate of $\tilde{\gamma}_k$ is in each iteration
MMV-AMP iteration from the
row vectors
\beq
\rv^k_{t} = (\Am^\herm\Zm^{t} + \Xm^{t})_{k,:}
\eeq
under the assumption that the decoupled channel model holds true, i.e.
that $\rv^k_t \sim \mathcal{CN}(\xv_k,\text{diag}(\tauv_t))$
and $\xv_k$ follows a conditional Gaussian distribution
$\xv_k \sim \mathcal{CN}(0,\tilde{\gamma}_k\Id_M)$, therefore
$\rv_t^k \sim \mathcal{CN}(0,\tilde{\gamma}_k\Id_M + \text{diag}(\tauv_t))$.
Then the ML estimate of $\tilde{\gamma}_k$ at iteration $t$ is given by
\beq
\hat{\gamma}_k^\text{ML} = \max\left(0,\frac{\|\rv_t^k\|_2^2}{M} - \frac{\sum_{m=1}^M \tau_m}{M}\right) 
\eeq 
We set 
\beq
\hat{g}_k = \max(\hat{\gamma}_k^\text{ML},g_\text{min})
\label{eq:gmin}
\eeq
to avoid too small estimates of $g_k$ for inactive
columns which would lead to undesired overfitting effects.
The denoiser is then chosen as
the conditional mean estimator \eqref{PME-cond} but with the true
value of $\tilde{g}_k$ replaced by its regularized ML estimate:
\beq
\eta_k^\text{ML}(\rv,\tauv) = \eta_k(\rv,\hat{g}_k,\tauv)
\eeq
The conditional PME (\ref{PME-cond}) has been given in \cite{Liu2018a}, we recall the equations here
for completeness:
\beq
\eta_{k}(\rv,g_k,\tauv) = \phi_{k}(\rv,g_k,\tauv)g_k(g_k\Id_M + \text{diag}(\tauv))^{-1}\rv,
\label{eq:eta}
\eeq
where the coefficient $\phi_{k}(\rv,g_k,\tauv)\in[0,1]$
is given by
\beq
\begin{split}
    &\phi_{k}(\rv,g_k,\tauv) \\
    &= \left\{1 + \frac{1-\lambda}{\lambda}  \prod_{i=1}^M \left [ \frac{g_k+\tau_{i}^2}{\tau^2_{i}} \exp\left ( - \frac{g_k |r_i|^2}{\tau_{i}^2(g_k + \tau^2_{i})}\right)\right ] \right\}^{-1}
    \label{eq:vamp_pme_b}
\end{split}
\eeq
with $\lambda = \frac{K_a}{N}$.

The MMV-AMP iterations are repeated $T$ times, then the active set can be obtained in two ways.
The first it to pick the indices with the
$K_a + \Delta$ largest $\hat{\gamma}_k$, where $\Delta$ is a
parameter which allows to control the number of false alarms. While this method
performs generally well it requires an estimate of $K_a$.
Alternatively one can pick a sequence of thresholds $\theta_1,...,\theta_N$ and define
\beq
\mathcal{A} = \{k:\|\rv^k_{T}\|_2^2 > \theta_k\}.
\label{eq:thresholding}
\eeq
The thresholds can be chosen as
\beq
\hat{\theta}_k = M\frac{\tau_T^2(\hat{g}_k + \tau_T^2)}{\hat{g}_k}\log\left(1 + \frac{\hat{g}_k}{\tau_T^2}\right)
\label{eq:thresh_ml}
\eeq 
which coincide with the thresholds suggested in \cite{Liu2018a} only that the true values of $\tilde{g}_k$
are replaced by their ML estimates. This choice of thresholds was motivated in \cite{Liu2018a}
by the fact that
they leads to vanishing activity detection error probabilities in the limit $M\to\infty$.
The activity detection error probabilities $P_\text{md}^\text{AD}$ and $P_\text{fa}^\text{AD}$
are defined as the probabilities of missing an active column of $\Am$ or falsely declaring
an column as a active ,
not to be confused with the URA per-user error probabilities
defined in \eqref{eq:ura_pmd} and \eqref{eq:ura_pfa}.
\subsection{Posterior Mean Estimation}
\label{sec:amp_pme}
A more practical version of the posterior mean estimator can be obtained when the received power
is limited to be in a discrete set $(\pi_1,...,\pi_G)$.
This is the case when the partial SCI policy described in Section \ref{sec:power_control}
is employed where the probability of a user
being received on level $i$ is given by $\xi_i$. Furthermore, a discrete distribution can be used to
approximate a continuous distribution $p_g$. In that case we set $\xi_i = p_g(\pi_{i-1} < P_kg_k \leq \pi_i)$.
In both cases we employ 
\eqref{eq:su_approx} and ignore the possibility of collisions, since they would lead to intractable equations.
With \eqref{eq:su_approx} the posterior probabilities of the power levels in the decoupled
Gaussian channel are given by:
\beq
\begin{split}
p(\pi_i|\rv) &= \frac{p(\rv|\pi_i)p(\pi_i)}{p(\rv_k)} \\
             &= \frac{\xi_i}{p(\rv_k)}
             \left[\frac{\lambda}{\pi_i + \tau^2}\exp\left(-\frac{\|\rv_k\|_2^2}{\pi_i + \tau^2}\right)\right.\\
             &\quad +\left.\frac{1-\lambda}{\tau^2}\exp\left(-\frac{\|\rv_k\|_2^2}{\tau^2}\right)\right]
\end{split}
\eeq
where $p(\rv_k) = \sum_{i=1}^{G}p(\rv_k|\pi_i)p(\pi_i)$.
With this, the posterior mean estimator \eqref{eq:uncond_eta} takes on the form:
\beq
\begin{split}
&\eta_k(\rv_k,\tauv)\\
&= \left[\sum_{i=1}^Gp(\pi_i|\rv)\phi_{t,k}(\rv,\pi_i,\tauv)\pi_i(\pi_i\Id_M + \text{diag}(\tauv))^{-1}\right]\rv_k
\end{split}
\eeq 
After the final iteration of MMV-AMP the posterior distribution can be used to find the MAP estimate
\beq
\hat{\pi}_k = \argmax_{\pi \in (\pi_1,...,\pi_G)}p(\pi|\rv_k).
\eeq

The ML estimation method from the previous
section can of course still be used with discrete power levels
since it treats the LSFCs as deterministic unknowns. In addition, the estimate can be improved by rounding the final 
ML estimates to the nearest allowed power level. AD can then be done as described in 
Section \ref{sec:lsfc_est}.
\subsection{Analysis}

\subsubsection{MRC Analysis}
\label{sec:mrc_analysis}
In this section we calculate an approximate finite-blocklength lower bound on the error probability and
on the energy efficiency of the MRC approach. 
For the purpose of analysing the MRC step we assume that the identification of the active columns
can be done without errors.
This is justified empirically, since for 
for the range of parameters we use in the simulations section the observed error rates are
in the order of $10^{-5}$ or less.
Since this is much smaller than the targeted PUPEs we neglect them. 

In addition, we restrict the MRC analysis to the case where $\tilde{\Gammam}$ is known, i.e. $\hat{\Gammam} = \tilde{\Gammam}$.
The general case with mismatched LSFCs can be treated in a similar way, but leads to additional interference terms
which do not admit a simple expression. Also we ignore the possibility of collisions in the analysis since
the probability of a collision for the parameters of interest is much smaller than the targeted per-user error probability.

With this simplifications we obtain a lower bound
on the error probability
and 
in the regime where $K_a<n_p$ we expect it to be tight,
as in this regime the AD error rates and the error of the LSFC
estimation are very low \cite{Liu2018a, Fen2021a}. 
The covariance of the channel estimation error of the LMMSE estimation in \eqref{eq:lmmse} is given by
\beq
\Cm_e = \Id_{K_a} - \Gammam_\mathcal{I}^{1/2}\Am_\mathcal{I}^\herm \left(\Am_\mathcal{I}\Gammam_\mathcal{I}\Am_\mathcal{I}^\herm + N_0\Id_{n_p}\right)^{-1}\Am_\mathcal{I}\Gammam_\mathcal{I}^{1/2}
\label{eq:error_cov}
\eeq 
and the MSE of the channel estimate of user $k\in\mathcal{K}_a$ is given by
$\sigma^2_k := \EE\{|h_{k,m}-\hat{h}_{k,m}|^2\} = (\Cm_e)_{k,k}$.
A typically tight approximation of the effective $\SINR$ of each user after MRC  
is \cite{Mar2016,Cai2018}
\beq
\SINR_k = \frac{M(1 - \sigma_k^2) g_k P}{N_0 + \sigma_k^2 g_k P +  \sum_{j=1, j \neq k }^{K_a}   g_j P} 
\label{eq:sinr}
\eeq 
For orthogonal pilots the channel estimation error reduces to
\beq
\sigma^2_k = \frac{N_0}{N_0+n_p P g_k}    
\label{eq:ortho}
\eeq 
which lower bounds the actual channel estimation error. 
Although \eqref{eq:ortho} is a crude bound, it is useful when the evaluation of $\eqref{eq:error_cov}$
is computationally too expensive since it requires the inversion of a possibly large matrix. 
We use it only in the optimization procedure in Section \ref{sec:opt}

An approximation of the achievable rates of a block-code with block-length $2n_d$ and error probability $p_e$
on a real AWGN channel with power $\SINR$ is given by the normal approximation \cite{Pol2010}
\beq
    R \approx 0.5\log(1+\SINR) - \sqrt{\frac{V}{2n_d}}Q^{-1}(p_e)
    \label{eq:normal}
\eeq 
where 
\beq
V = \frac{\SINR}{2}\frac{\SINR+2}{(\SINR+1)^2}\log^2 e
\eeq
and $Q(\cdot)$ is the Q-function. Using the normal approximation we can find the required $\SINR$ to achieve
a certain error probability at a given block-length and then we can find the required input power
to achieve the target \SINR. 

\begin{figure}
   \centering
   \includegraphics[width=\linewidth]{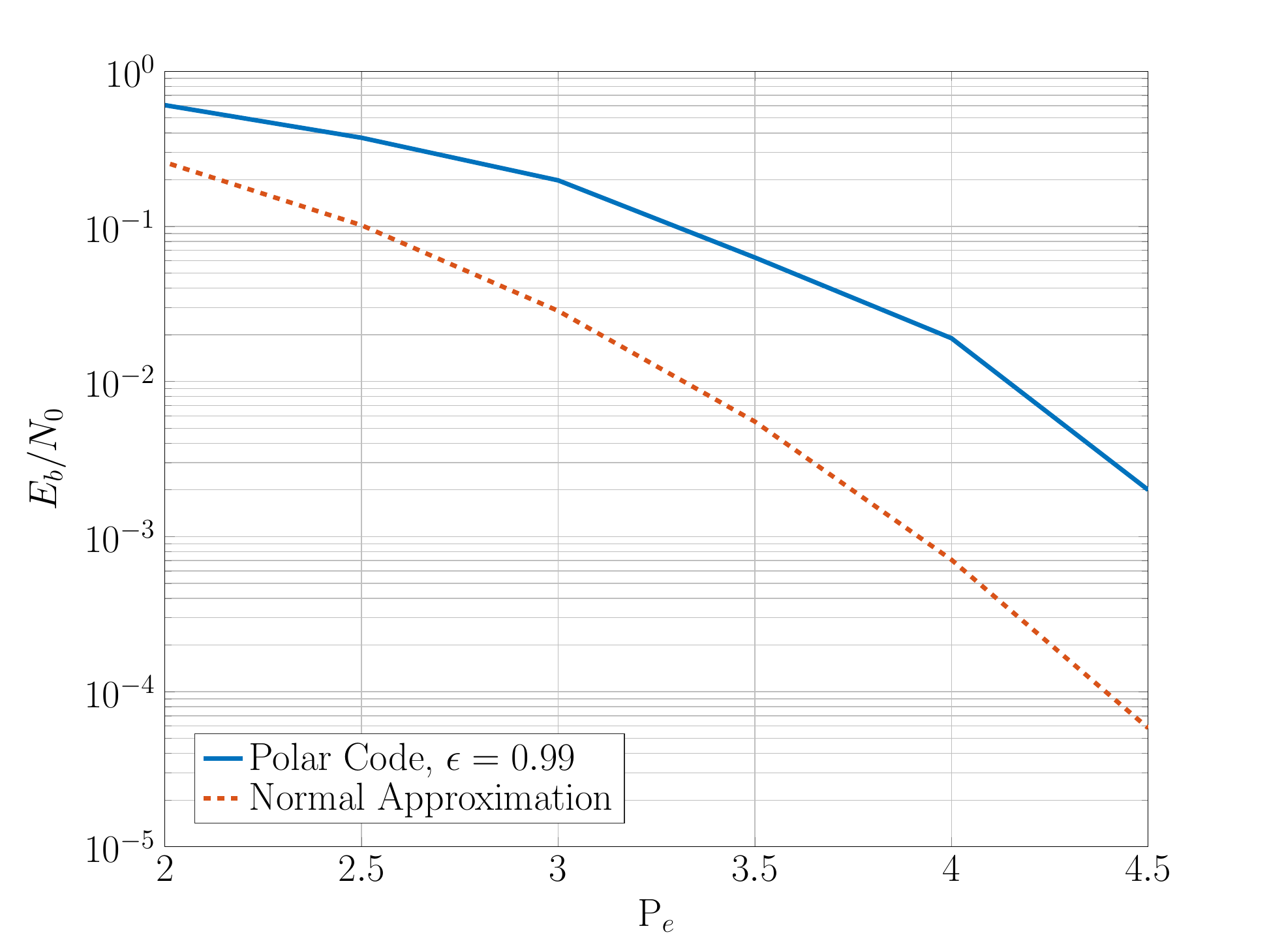}
   \caption[$p_e$ over $E_b/N_0$]{Polar code with $n=4096$ and $84$ bits vs normal approximation.} 
   \label{fig:pe_ebn0_pc}
\end{figure}

\begin{remark}
The normal approximation here is used as an easy-to-evaluate formula for the achievable
error probability in a Gaussian channel at some $\SINR$ and $n_d$. If the goal is to analyse a given
single-user code, much more accurate results can be obtain by empirically obtaining the
$p_e$ over $\SNR$ relation for this code on an AWGN channel
and using this relation instead of the normal approximation, see \figref{fig:pe_ebn0_pc}.
To better estimate the best achievable error probabilities tighter formulas than
the normal approximation
are available \cite{Pol2010,Ost2021}.
\end{remark}

\subsection{SIC}
Under full SIC, i.e. each message is individually recovered and subtracted from $\Ym_d$,
starting with the user with the largest
LSFC (assuming $g_1 > g_2 > ...$), the average effective SINRs are given by
\beq
\begin{split}
               \SINR_k
               = \frac{M(1-\sigma_k^2)g_kP}{N_0 + \sum_{i=1}^{k-1}I^R_i + \sum_{j=k+1}^{K_a}g_jP}
\label{eq:sinr_sic}
\end{split}
\eeq 
where
\beq
\text{I}^R_i = (1-\epsilon_i)\sigma_i^2g_iP + \epsilon_i g_iP
\eeq
is the residual interference left after interference cancellation of user $i$.
$\epsilon_k \in \{0,1\}$ is a Bernoulli random variable indicating whether the codeword of user $k$
has been erroneously
decoded or not.
$P(\epsilon_k = 1) = p_e(\SINR_k)$ can be calculated from $\SINR_k$ in sequential order, $k=1,...,K_a$ by the normal approximation \eqref{eq:normal}
or the $p_e$-over-$\SNR$ curve for a given code.
Note, that even if a message is decoded correctly, i.e. $(\epsilon_k = 0)$, a residual
interference $\sigma_k^2g_kP$ remains due to the channel estimation error.
The vector $\SINR = (\SINR_1,...,\SINR_{K_a})$ is random due to the presence of 
the $\epsilon_k$. For a target outage probability $\delta$ we define a vector $\underline{\SINR} = (\underline{\SINR}_1,...,\underline{\SINR}_{K_a})$
by the property
\beq
    p(\SINR_k < \underline{\SINR}_k) = \delta \quad \forall k,
\eeq 
where the probability is over $\epsilon_k$ and $g_k$. With this an upper bound on $p_\text{md}$
is given by
\beq
p_\text{md} \leq \frac{1}{K_a}\sum_{k=1}^{K_a}\left[(1-\delta)p_e(\underline{\SINR}_k) + \delta\right].
\eeq
$\underline{\SINR}$ can be numerically estimated by sampling $N_s = \mathcal{O}(\delta^{-1})$ vectors of $\SINR$s and
then choosing the $\lfloor N_s\delta \rfloor$'th smallest entry of $(\SINR_k^1,...,\SINR_k^{N_s})$
as $\underline{\SINR}_k$.

Although full SIC leads to the best SINRs it is computationally
inefficient since single-user decoding can not be done in parallel
which makes it not practicable for a large number
of active users.
Therefore, we consider a simplified SIC scheme where, in the MRC phase, the BS divides the active users into
$G$ groups, based on their received power. 
Then all messages within one group are decoded and subtracted in parallel
starting from the group with the highest average power.
Let $(\mathcal{G}_q)_{q=1,...,G}$ be some partition of $[1:K_a]$ into $G$ groups,
i.e. $\bigcup_{q=1}^{G}\mathcal{G}_q = [1:K_a]$ 
and $\mathcal{G}_q \cap \mathcal{G}_l = \emptyset$ for $q\neq l$. Furthermore, assume that the 
groups are fully ordered according to the LSFCs, i.e. $g_i \leq g_j$ for all $i\in\mathcal{G}_q,j\in\mathcal{G}_l$
if $q \geq l$.
Let $\SINR_k^q$ denote the $\SINR$ of user $k$ in group $q$ after the contributions of groups $1,...,q-1$
have been cancelled.
They are given by:
\beq
\SINR_k^q = 
\frac{M(1-\sigma_k^2)g_kP}{N_0 + \sum_{i=1}^{q-1}\sum_{k\in\mathcal{G}_i}I_k^R + \sum_{j=q}^{G}\sum_{k\in\mathcal{G}_j}g_kP}
\label{eq:sinr_sic_grouped}
\eeq 

\begin{figure}
   \centering
   \includegraphics[width=\linewidth]{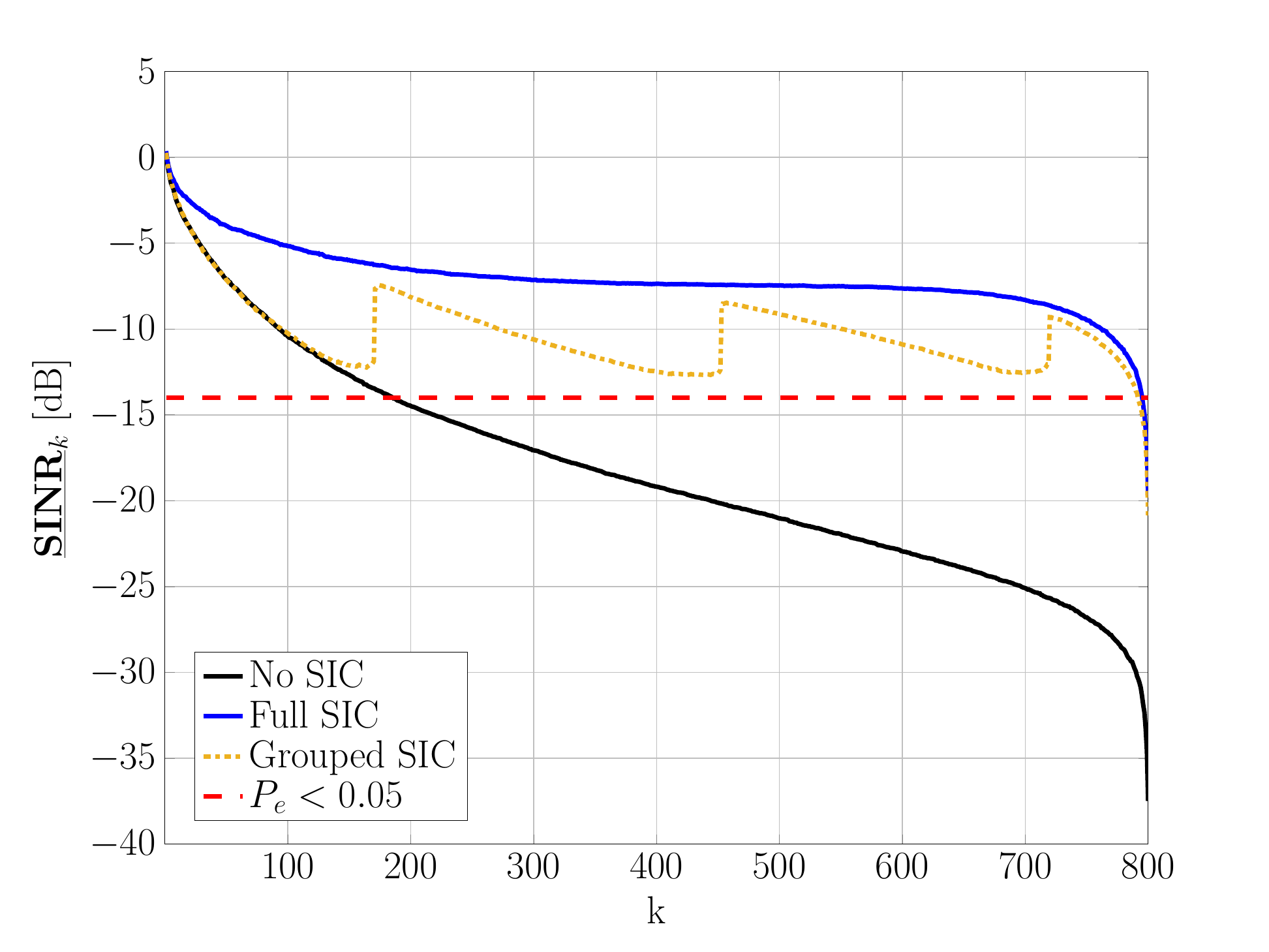}
   \caption[SINR SIC]{SINR profile under different SIC strategies. The dashed line represents the SINR necessary
   to achieve $P_e < 0.05$ according to the normal approximation.
   The LSFCs follow the shadowing-pathloss distribution \eqref{eq:shp}.} 
  \label{fig:SINR_SIC}
\end{figure}

In \figref{fig:SINR_SIC} we compare the $\underline{\SINR}$ profiles under
the different SIC strategies according to
formulas \eqref{eq:sinr},\eqref{eq:sinr_sic} and \eqref{eq:sinr_sic_grouped}
under the NPC policy with $R = 1$ km, $\alpha = 128.1,\beta = 36.7, g_\text{max} = -3$ dB.
We choose $K_a = 800, M=50, n=3200, L = 1152$.
For simplicity we used the approximation \eqref{eq:ortho} for the channel estimation errors. 
For the grouped approach we have divided users into four groups
$\mathcal{G}_i = \{k: o_{i-1} > g_k \geq o_i\}$
with dividing points $(o_0,..,o_4) = (\infty,-16,-23,-28,-\infty)$ dB.
$\delta$ and the dashed line in \figref{fig:SINR_SIC} are chosen such that
$(1-\delta)p_e(\underline{\SINR}_k) + \delta < 0.05$
when
$\underline{\SINR}_{k}$ is above the dashed line.
We can see that without SIC less than $150$ users have an acceptable SINR. 
Furthermore, the picture shows that four groups are sufficient to achieve
an error probability similar to full SIC.

Formula \eqref{eq:sinr_sic_grouped} simplifies when in addition to grouped SIC
the partial SCI policy of Section \ref{sec:power_control} is employed. For this we 
assume again that the active users are partitioned into $\mathcal{G}_1,...,\mathcal{G}_G$ and furthermore that
each user in $\mathcal{G}_q$ is received with the same power $\pi_q$, satisfying $\pi_1 > \pi_2 > ... >\pi_G$.
Let $n_q = |\mathcal{G}_q|$ denote the number of users in group $q$.
Then each user in group $q$ has the same SINR given by:
\beq
\SINR^q = \frac{M(1-\sigma_q^2)\pi_q}{N_0 + \sum_{i=1}^{q-1}[(n_i - \zeta_i)\sigma_i^2 \pi_i + \zeta_i \pi_i] + \sum_{j=q}^{G}n_j \pi_j}
\label{eq:sinr_sic_pc}
\eeq 
where $\zeta_q = \sum_{k\in\mathcal{G}_q}\epsilon_k$ is the number of wrongly decoded messages in group $q$.
Note, that in the partial power control policy users chooses their own group based on their LSFC which has been estimated
in the downlink. Therefore, $n_q$ is in general a random variable.

When the group size are large enough,
since users choose their group independent of each other, the distribution of $(n_1,...,n_G)$
will concentrate sharply around its
mean $\EE (n_1,...,n_G) = K_a(\xi_1,...,\xi_G)$ where $\xi_q$ is the probability that a user chooses group $q$.
For a given distribution of LSFCs these are given as $\xi_q = \PP(o_{q-1} < g_k \leq o_q)$.
The concentration can be more precisely stated as a large-deviation property \cite[Thm. 2.4.3]{Ell2006}:
\beq
\lim_{K_a\to\infty}\frac{1}{K_a}\log \PP(n_1,...,n_G) = D_{KL}\left(\frac{n_1}{K_a},...,\frac{n_G}{K_a}\middle\|\xi_1,...,\xi_G\right)
\label{eq:large_dev}
\eeq 
which implies exponential convergence of $(n_1,...,n_G)$ to its mean with $K_a$. Therefore, if the
group sizes are sufficiently large we can approximate the random variables $n_i$ by $K_a\xi_i$.
The same holds for the RVs $\zeta_i$ which are also sums of iid RVs.
With this approximation the $\SINR$ profile is not a random variable anymore but depends only
on the power levels and their occupancies. In the following section we propose methods
to optimize
those parameters.
\subsection{Optimization of the power levels}
\label{sec:opt}
\begin{figure}
   \centering
   \includegraphics[width=\linewidth]{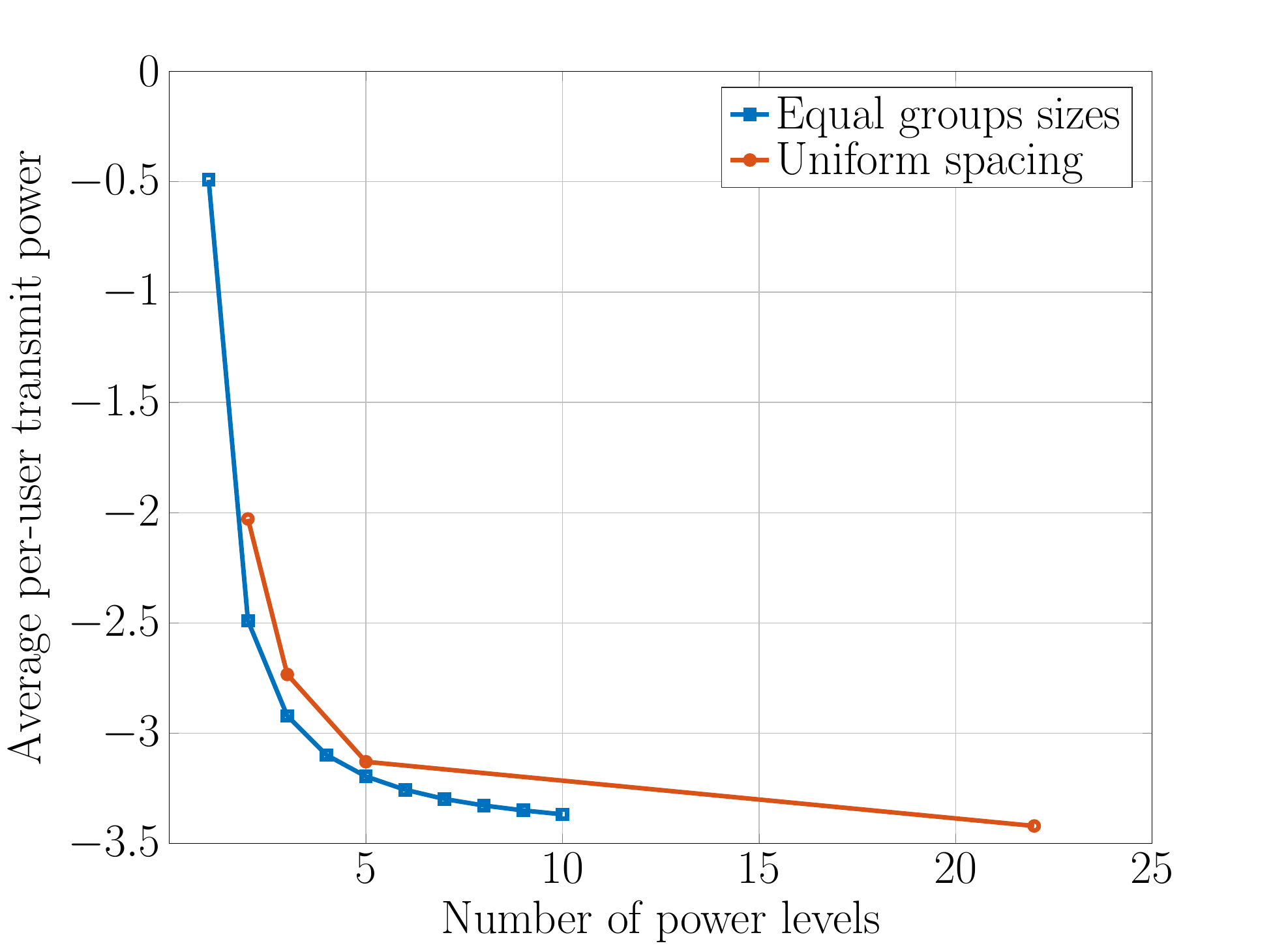}
   \caption[SINR SIC]{Gains of the transmit power \eqref{eq:transmit_power} in dB by introducing
       multiple received power levels according to formula \eqref{eq:sinr_sic_pc}.
   Power levels are obtained by the two optimization methods introduced in \ref{sec:opt}. The LSFCs follow
   the shadowing-pathloss distribution \eqref{eq:shp} and $K_a = 800$. The remaining parameters 
   are chosen as in Section \ref{sec:simulations}.} 
   \label{fig:transmit_gains}
\end{figure}

Let the power control policy be defined by a set of corner points $(o_1,...,o_G)$ and a set of power levels
$(\pi_1,...,\pi_G)$.
We wish to optimize these parameters w.r.t. the average transmit power
\beq
P_T = \EE\left[\sum_{k=1}^{K_a}\pi_{q(k)}g_k^{-1}\right] = \sum_{i=1}^{G}\pi_{i}\EE\left[\sum_{k\in\mathcal{G}_i}g_k^{-1}\right],
\label{eq:transmit_power}
\eeq
where $q(k)$ denotes the group of user $k$,
under the constraint that the average PUPE remains below $P_e$.
Given the recursive nature of \eqref{eq:sinr_sic_pc} this
is a highly non-convex problem.
We introduce two solutions based on different approximations and compare them.\\
1) Equal group size constraint\\
In this approach we fix the number of groups $G$ and let each group have the same expected size,
i.e. the corner points are chosen such that $\xi_i = K_a/G$. This allows to calculate the coefficients
$\EE\left[\sum_{k\in\mathcal{G}_i}g_k^{-1}\right]$ and minimize the transmit power \eqref{eq:transmit_power} w.r.t. 
$(\pi_1,...,\pi_G)$ under the constraint $\SINR^q \geq \SINR^*$.
\begin{subequations}
\begin{alignat}{2}
&\!\min_{(\pi_1,...,\pi_G)}        &\qquad& \sum_{q=1}^{G}\pi_{q}\EE\left[\sum_{k\in\mathcal{G}_q}g_k^{-1}\right] \label{eq:optProb}\\
&\text{subject to} &      &\forall q: \SINR^q \geq \SINR^*,\label{eq:constraint1}
\end{alignat}
\label{eq:opt_equal}
\end{subequations}
2) (Almost) Linear optimization \\
In this approach we introduced an overcomplete set of $W$ power levels. E.g. all equally spaced power levels
between -30 and 0 dB. Then the optimization problem is formulated to find the optimal fraction of active users
for each power level. Since the coefficients $\EE\left[\sum_{k\in\mathcal{G}_q}g_k^{-1}\right]$ 
in the transmit power \eqref{eq:transmit_power} depend on the group sizes they cannot be precomputed and make the 
objective function very complicated. Therefore, we use only the total received power as objective function,
given by $P_R = \sum_{q=1}^W\xi_q\pi_q$ leading to the following optimization problem.
\begin{subequations}
\begin{alignat}{2}
&\!\min_{\xi_1,...,\xi_W} &\qquad& \sum_{q=1}^{W}\xi_{q}\pi_{q} \label{eq:optProb2}\\
&\text{subject to} &      &\forall q: \SINR^q \geq \SINR^*, \quad \text{if } \xi_q > 0\label{eq:constraint21}\\
&                  &      & \sum_{q=1}^{W}\xi_q = 1.\label{eq:constraint22}
\end{alignat}
\label{eq:opt_linear}
\end{subequations}
Although there is no explicit constraint to keep the number of power levels with $\xi_q > 0$ low
the number of non-zero terms in the optimal solutions is usually much smaller than $W$.
We then let $G = |\{q:\xi_q > 0\}|$.
When a solution is found, the corner points, which define the groups assignments,
are chosen to get the resulting $\xi_q$. \\
Note, that neither of the two methods is a convex problem. The addition ``if $\xi_q >0$'' in
\eqref{eq:constraint21} makes the second method non-linear. Nonetheless,
common interior-point based non-linear solvers, like MATLABs $\mathsf{fmincon}$,
are able to solve those problems reliably.
The two approaches are compared in \figref{fig:transmit_gains}.
In approach 1 the number of levels is the only parameter as
opposed to approach 2 where the number of levels is an outcome of the optimization. For approach 2
we choose linearly spaced power levels with different values of the spacing. The optimal solution
puts weight only on few power levels the number of which is
summarized in Table \ref{table1}. For each spacing values the offset is optimized empirically over 5 values.
\begin{table}[h]
    \centering
    \begin{tabular}{|c|c|c|c|c|}
        \hline
        spacing [dB] & 2 & 1 & 0.5 & 0.1 \\
        \hline
        levels & 2 & 3 & 5 & 22 \\ 
        \hline
    \end{tabular}
    \caption{Number of power levels with $\xi_i > 0$ in the solution of \eqref{eq:opt_linear} depending on the spacing
    $\pi_{i} - \pi_{i+1}$.}
    \label{table1}
\end{table}

\section{Simulations}
\label{sec:simulations}
\subsection{Active Column Detection}

\begin{figure}
   \centering
   \subfloat[$P=-15$ dB]{\includegraphics[width=0.45\linewidth]{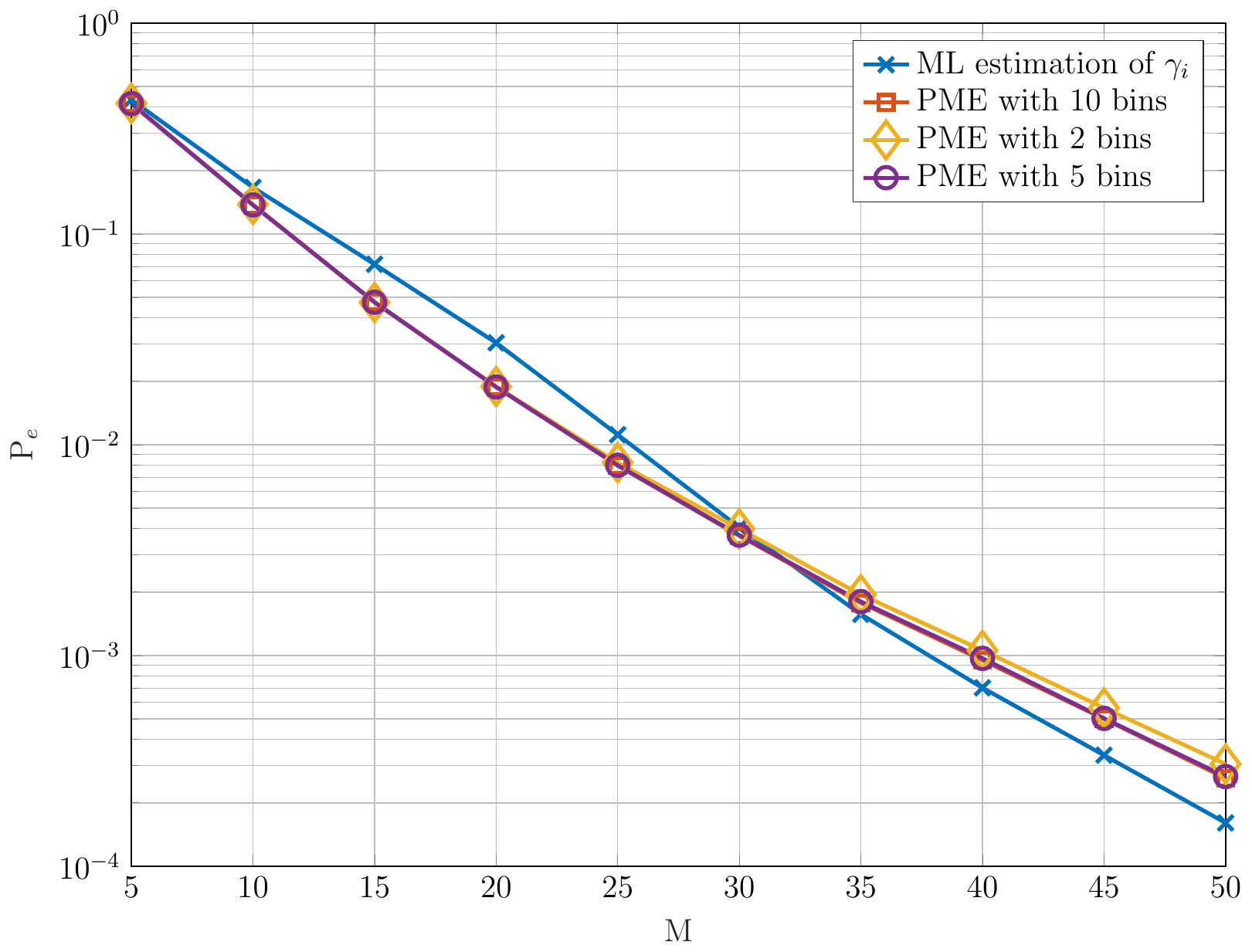}}\quad
   \subfloat[$M=10$]{\includegraphics[width=0.45\linewidth]{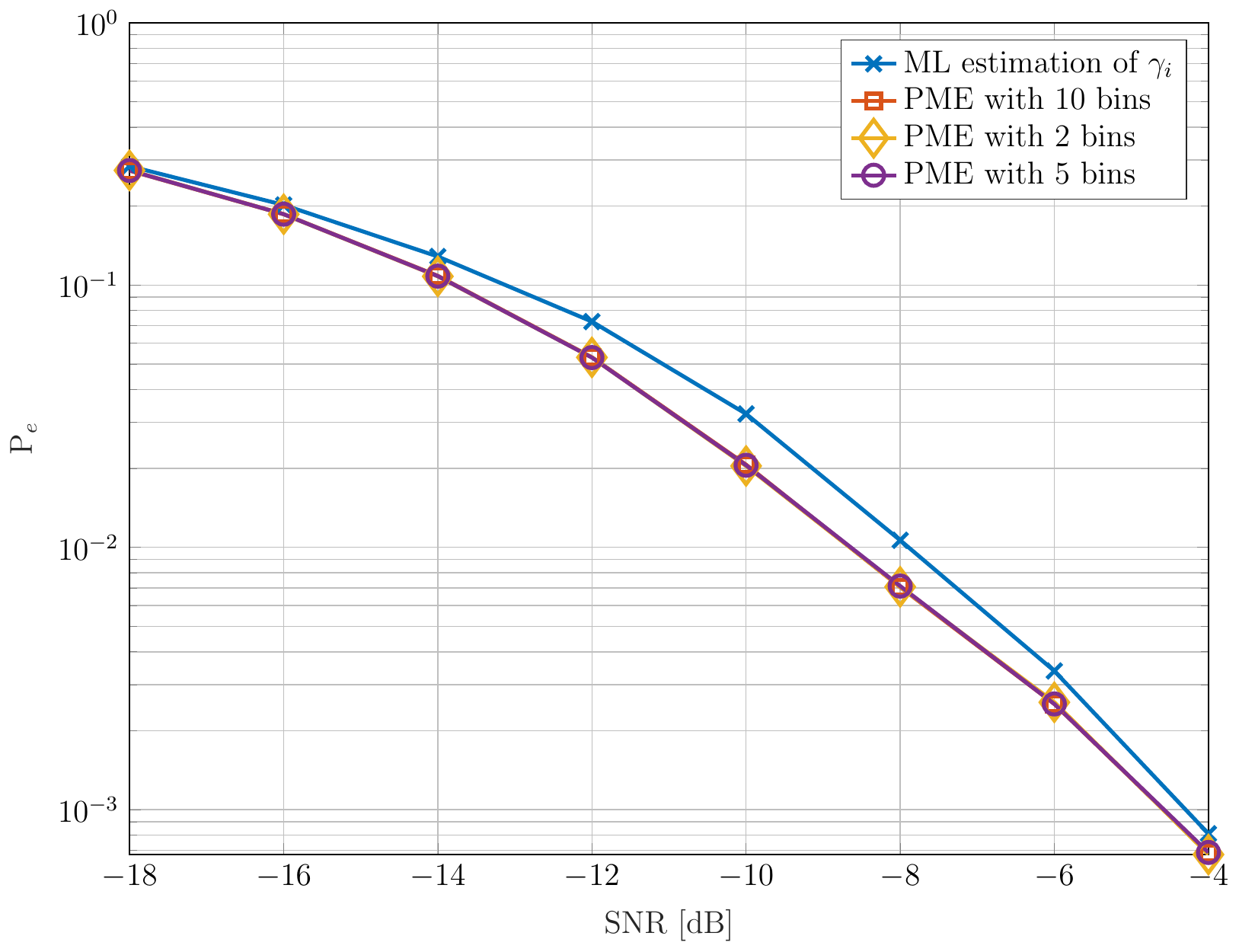}}\quad
   \caption[Titel des Bildes]{$P^\text{AD}_\text{md}$ under imperfect SCI
       with LSFCs distributed uniformly between $[-3,3]$ dB.}
   \label{fig:pe_uniform}
\end{figure}

\begin{figure}
   \centering
   \subfloat[$P = 18$ dB]{\includegraphics[width=0.45\linewidth]{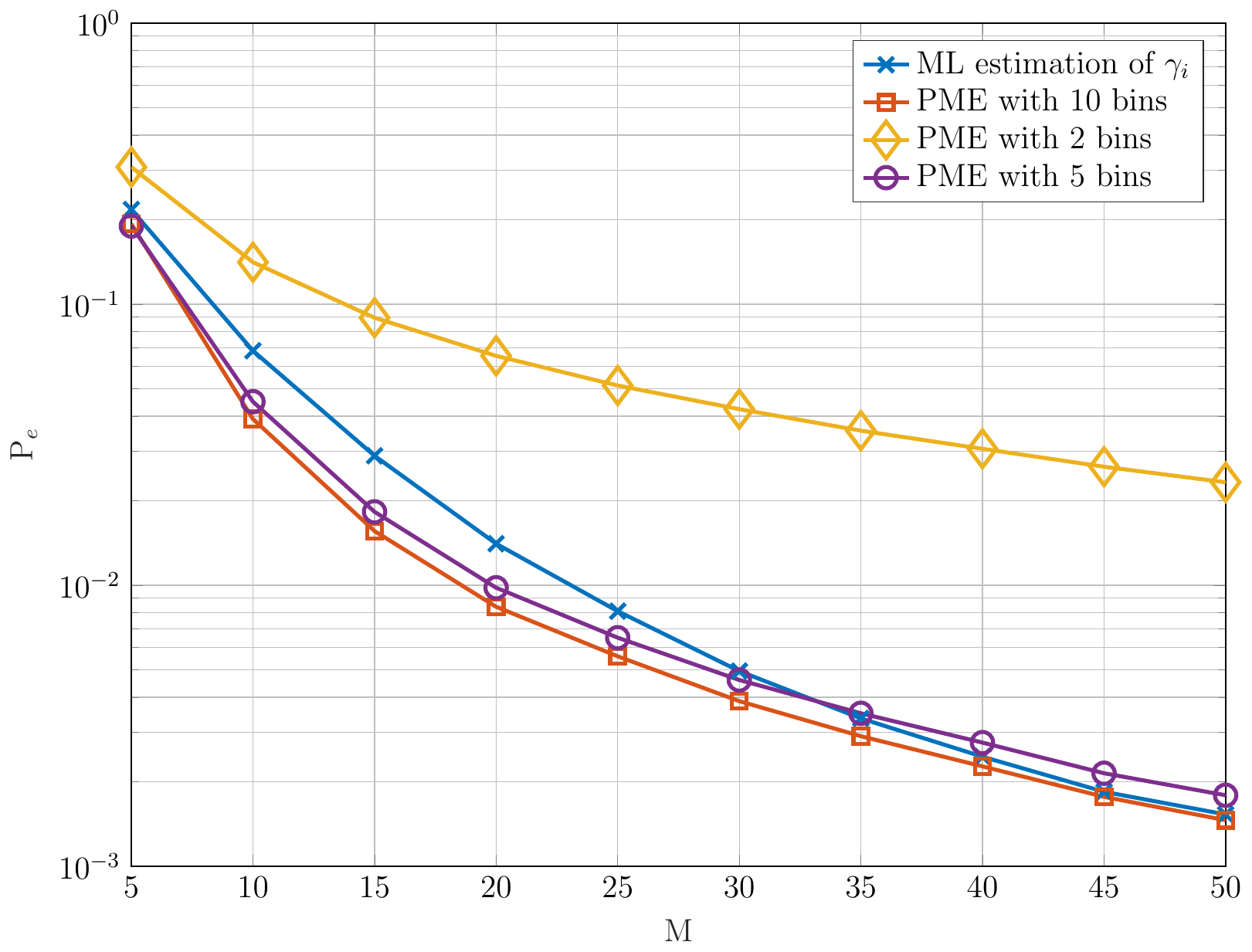}}\quad
   \subfloat[$M=15$]{\includegraphics[width=0.45\linewidth]{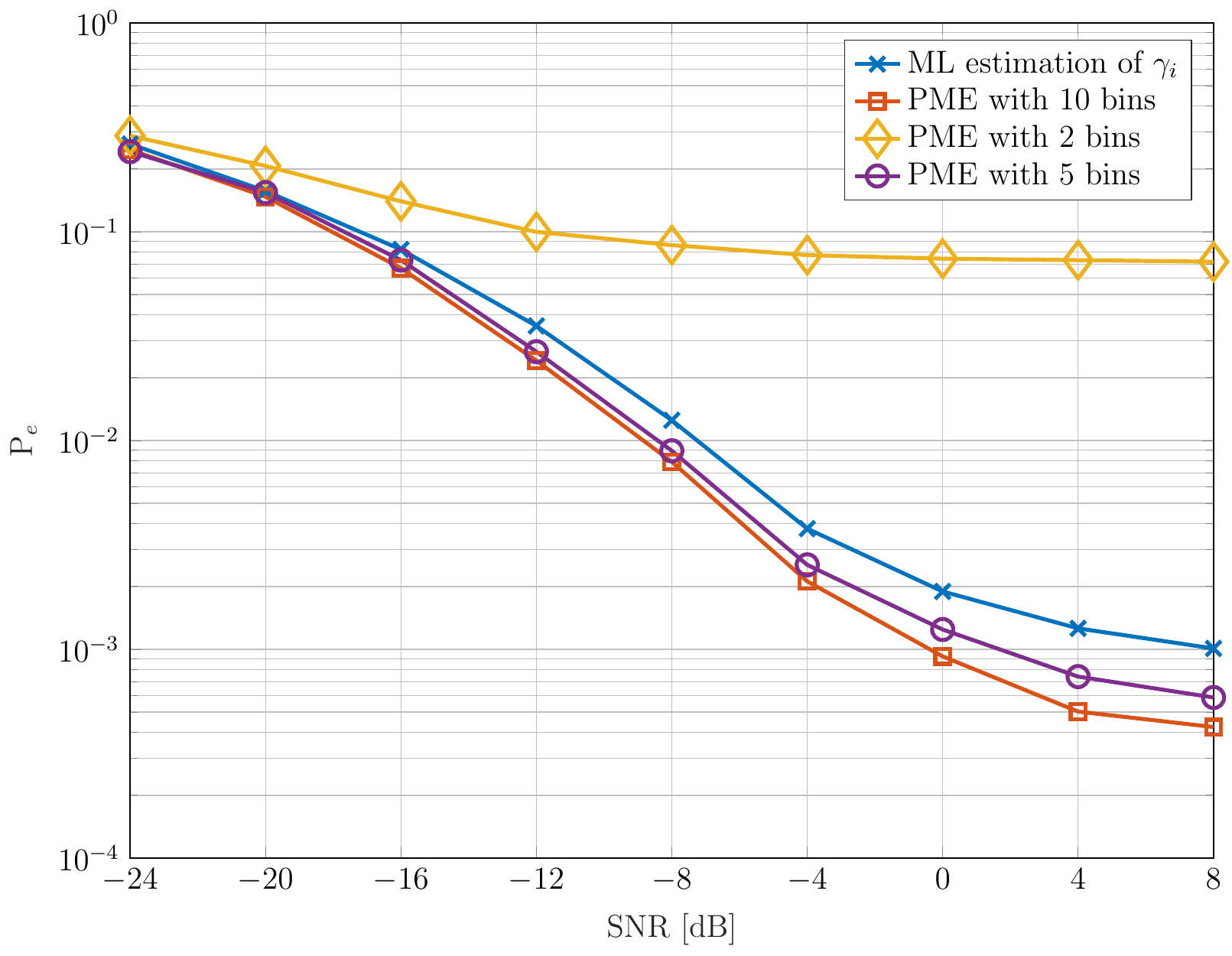}}\quad
   \caption[Titel des Bildes]{$P^\text{AD}_\text{md}$ under the NPC policy with LSFCs distributed according to the shadowing-pathloss model \eqref{eq:shp}.}
   \label{fig:pe_shp}
\end{figure}

\begin{figure}
   \centering
   \subfloat[imperfect SCI]{\includegraphics[width=0.45\linewidth]{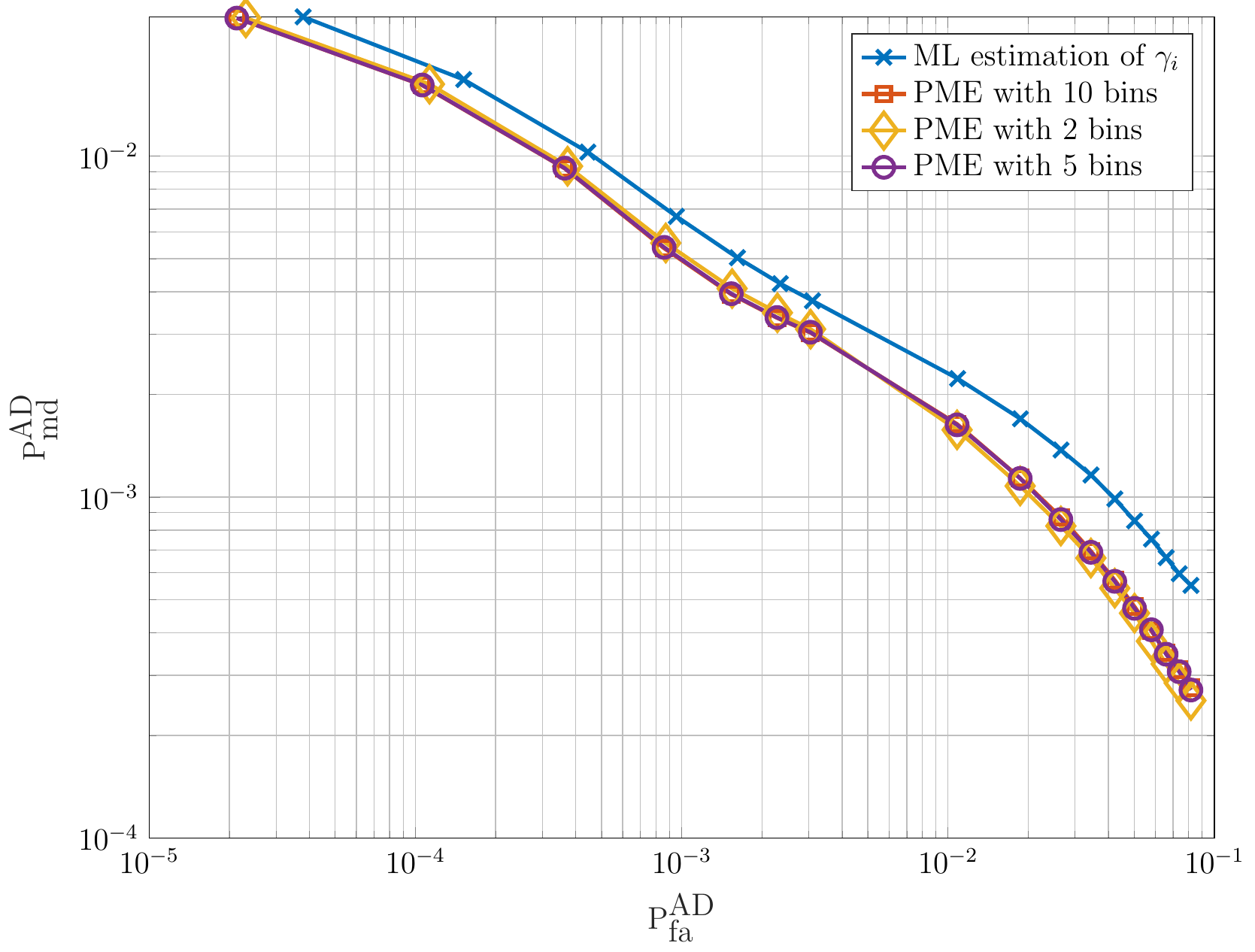}}\quad
   \subfloat[NPC]{\includegraphics[width=0.45\linewidth]{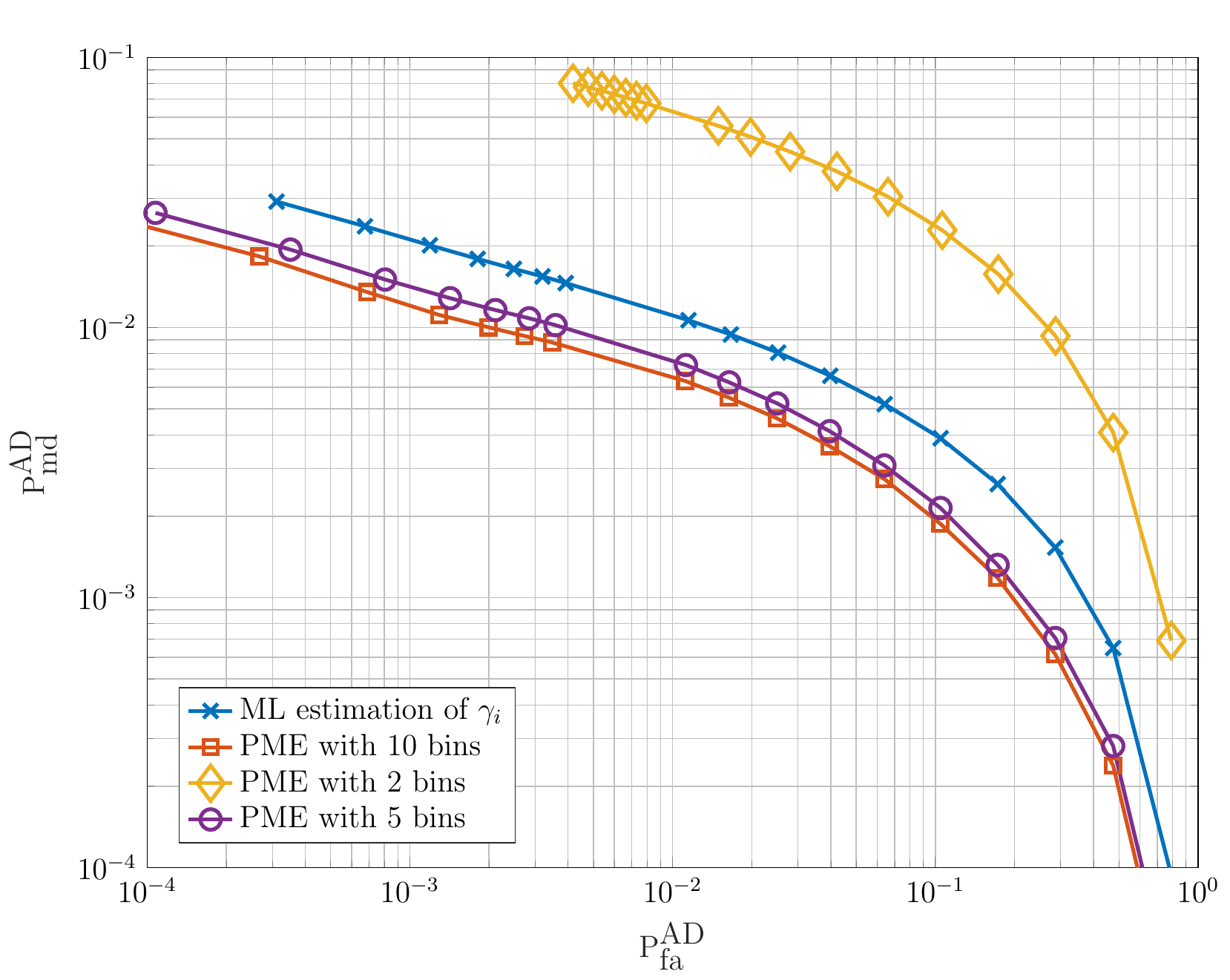}}\quad
   \caption[Titel des Bildes]{$P^\text{AD}_\text{md}$
   over $P^\text{AD}_\text{fa}$.
   The tradeoff is achieved by varying $\Delta$ as described in Section \ref{sec:lsfc_est}. Left: $M=15, P = -10$ dB. Right: $M=20, P= 18$ dB.}
   \label{fig:pe_roc}
\end{figure}

In this section we provide simulations of AD error probabilities
for MMV-AMP with the proposed ML estimation of LSFCs in Section \ref{sec:lsfc_est}.
For the simulations we choose $L=300, K_a = 300, N=2^{12}$. AD is done by picking the $K_a + \Delta$
columns with the largest $\hat{\gamma}_i$. In \figref{fig:pe_uniform} and \figref{fig:pe_shp} we
set $\Delta = 0$. In \figref{fig:pe_roc} $\Delta$ is varied between $\lfloor -0.06K_a \rfloor$ and
$ K_a$ to tradeoff between missed detections and false alarms.
We compare the ML estimation approach to the PME approach with a finite number of levels
as described in Section \ref{sec:amp_pme}. 
The levels and their distribution are obtained by approximating the distribution of LSFCs with
a histogram with a varying number of bins. The bins are chosen to be of equal size in log-scale.
\figref{fig:pe_uniform} shows $P^\text{AD}_\text{md}$ under imperfect SCI power control with
received powers distributed uniformly between -3 and 3 dB.
The parameter $g_\text{min}$ for the ML estimation of LSFCs in \eqref{eq:gmin} is chosen as
$g_\text{min} = -3$ dB, the lower limit of possible values.
We can see that the ML approach is close to the approximated PME and even outperforms it for large $M$.
Besides, we can observe that the performance of the approximate PME saturates at two levels.
A similar trend is observed with the more complicated RSS distribution under the NPC policy, only that
the approximate PME requires at least 5 levels to achieve an acceptable performance.
In \figref{fig:pe_shp} the NPC policy is used with 
$R = 1, r=0.25, \alpha = 128.1, \beta = 36.7, \sigma^2_\text{shadow} = 4$.
Limiting the largest possible received power by setting $r=0.25$
is crucial here because individual LSFCs which are much larger
than the other may
severely disrupt the convergence of the algorithm. 
We choose $g_\text{min} = 2\tau_T^2$ as parameter for the ML estimation,
where $T$ is the final iteration index.
Due to the shadowing component the LSFCs do no have a lower limit, therefore 
it seems more appropriate to choose $g_\text{min}$ relative to the noise variance in the decoupled
channel \eqref{eq:vamp_decoupled}.


\subsection{Unsourced Random Access}
\begin{figure}
   \centering
   \includegraphics[width=0.9\linewidth]{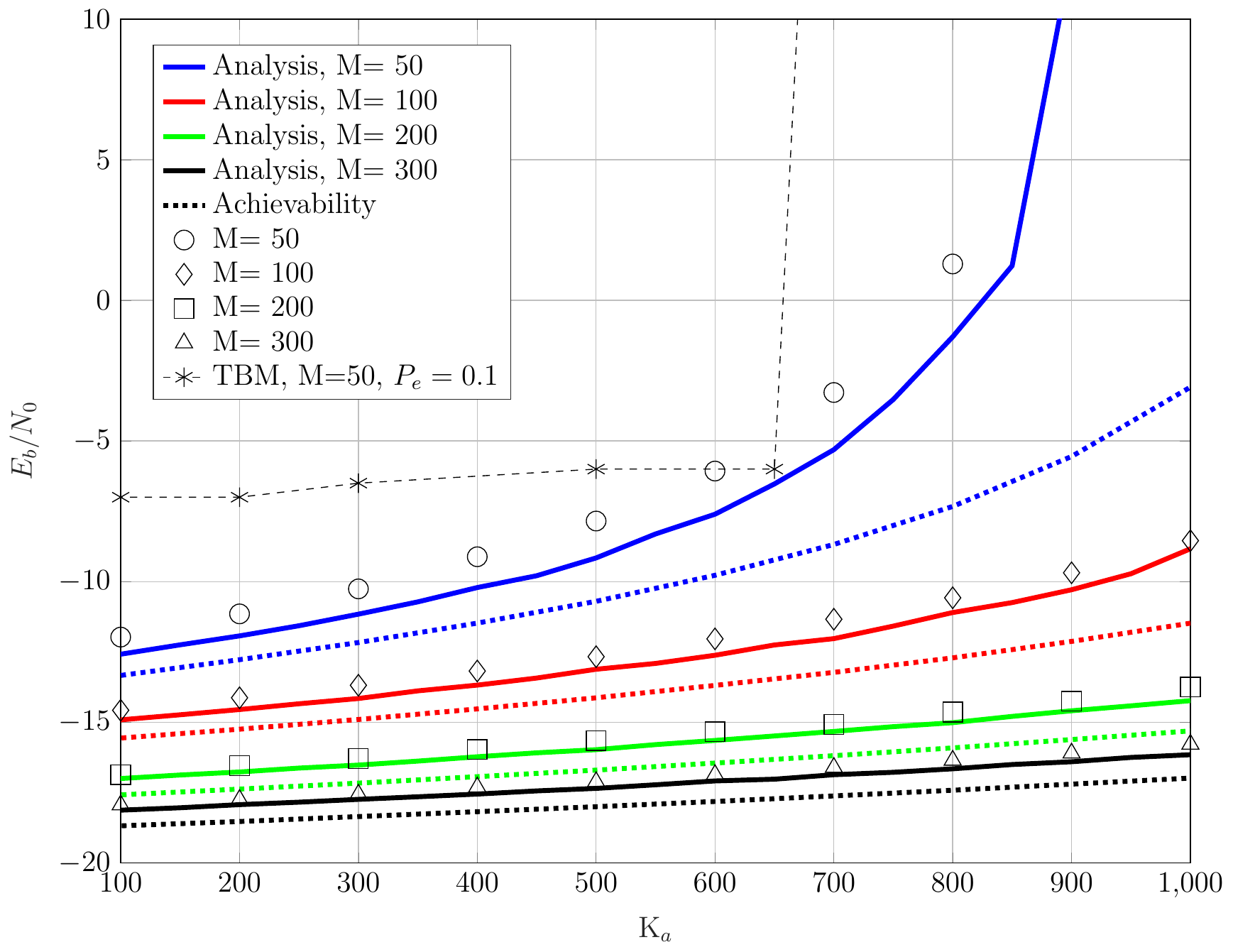}
   \caption[MRC: $E_b/N_0$ over $K_a$]{Required energy-per-bit with the MRC approach and perfect SCI to achieve
       $P_e = p_\text{md} + p_\text{fa} < 0.05$.
  Solid lines represent the theoretical estimates from Section \ref{sec:mrc_analysis} with the
  $P_e$ over $\SNR$ curve from the used polar code and DFT pilots. Dotted
  lines represent theoretical results using the normal approximation and Gaussian iid pilots.
  The markers represent simulation results.} 
  \label{fig:ebn0_over_K_mrc}
\end{figure}

\begin{figure}
   \centering
   \includegraphics[width=0.9\linewidth]{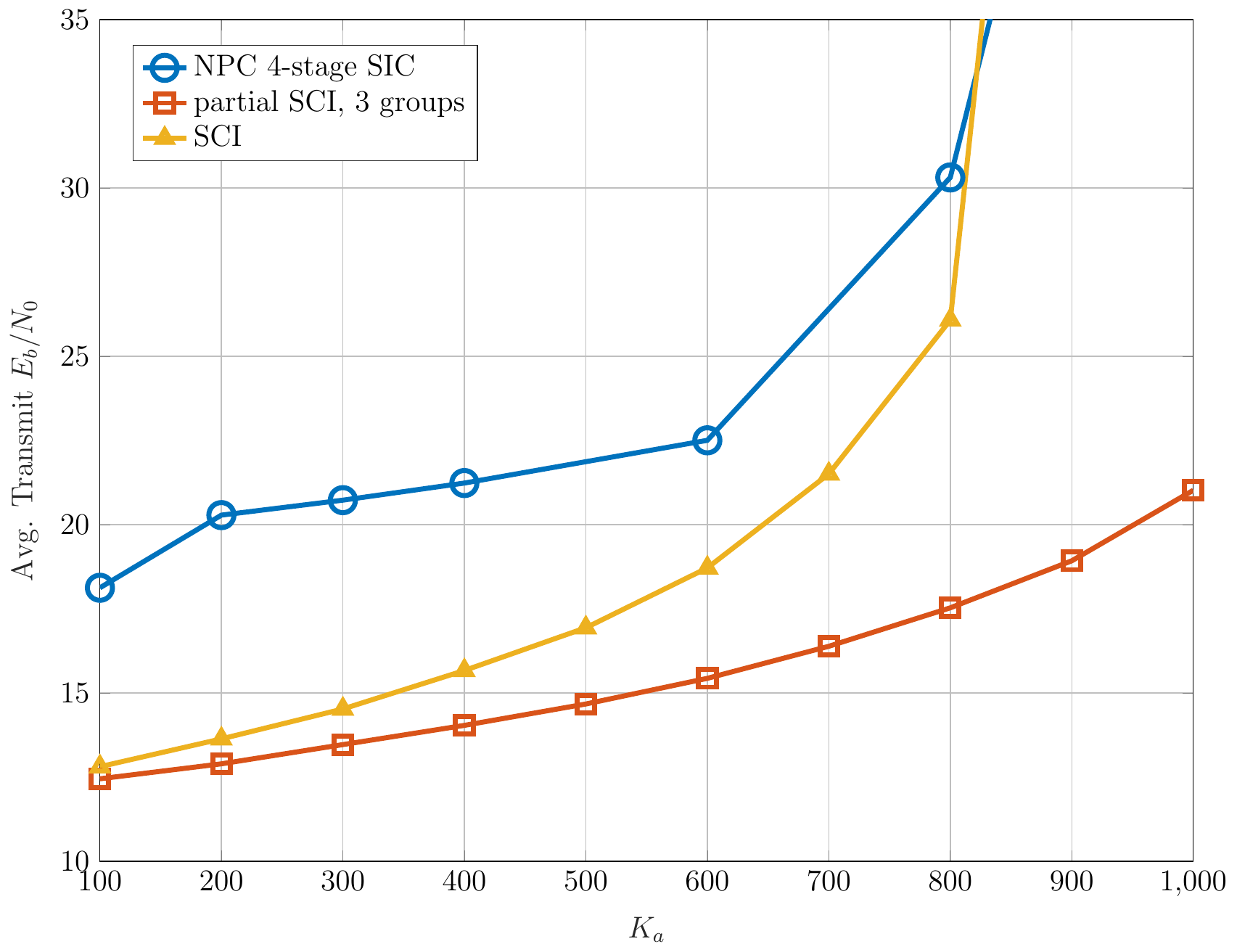}
   \caption[MRC: $E_b/N_0$ over $K_a$]{Average required transmit energy-per-bit
       with the MRC approach and $M=50$ to achieve
       $P_e = p_\text{md} + p_\text{fa} < 0.05$ under different power control policies and SIC strategies
       as described in the text.
       The LSFCs are assumed to follow the shadowing-pathloss distribution \eqref{eq:shp}.
       The partial SCI levels have been calculated according to the optimization \eqref{eq:opt_equal}.
       Note, that the levels are optimized only for $K_a = 800$.
  } 
  \label{fig:ebn0_over_K_sic}
\end{figure}

For the simulation in \figref{fig:ebn0_over_K_mrc} we choose $n=3200$, $P_e = 0.05$,
$B=100$, $n_p = 1152$, $n_d = 2048$
and $J = 16$.
A randomly sub-sampled DFT matrix is used as pilot matrix \cite{Fen2021e} and 
in the AD phase we use the MMV-AMP algorithm with an approximate calculation of the derivative
as described in Section \ref{sec:mmv_amp}.
For simplicity we assume that $K_a$ is known at the receiver, and after the MMV-AMP iterations
are finished the active columns
are estimated by picking the $K_a + \Delta$ indices with the largest estimated LSFCs.
We set $\Delta = \lfloor K_a/40 \rfloor$.
We use a polar code \cite{Ari2009,Bal2015} with an SCL decoder with
$16$ CRC bits and a list size of $32$. The polar code is constructed by the Bhattacharyya method
\cite{Tal2013a}.
\figref{fig:ebn0_over_K_mrc} shows the required $E_b/N_0$ to achieve
$P_e = p_\text{md} + p_\text{fa} \leq 0.05$ under a perfect SCI policy, i.e. all $g_k = 1$.
For comparison we add the reported values of the tensor-based-modulation (TBM) approach \cite{Dec2020c},
with tensor signature (8,5,5,4,4) and an outer BCH code,
although the values have been obtained with the higher value $P_e = 0.1$.
\figref{fig:ebn0_over_K_sic} shows the average transmit energy-per-bit $P_TB/(nK_a)$,
with $P_T$ defined in \eqref{eq:transmit_power},
under the shadowing-pathloss model \eqref{eq:shp} with parameters as above and $M=50$.
The SCI curve coincides with the $M=50$
curve in \figref{fig:ebn0_over_K_mrc} scaled by the average inverse channel coefficient
$\EE[\sum g_k^{-1}]/K_a$. For the NPC curve SIC is done in 4 groups defined by the LSFCs
division points $(-\infty,-26,-20,-12,\infty)$ dB. For the partial SCI policy we use three groups with
levels $(\pi_1,\pi_2,\pi_3) = (0,1,2.5)$ dB, which coincides, up to a common scaling factor and
some rounding errors, with the output of the optimization procedure
\eqref{eq:opt_equal} for $K_a = 800$. We can see that 
with only $3$ levels the average transmit energy can be reduced by about $3$ dB at $K_a = 800$
as predicted by \figref{fig:transmit_gains}. Furthermore, the grouped SIC allows to decode $K_a = 1000$
users at $M=50$ which was not possible without SIC. Note, that NPC does not work at all without
SIC and we have found empirically that 4-stage SIC is the minimum amount of stages required 
to decode up to $K_a = 800$ active users.
\subsection{Complexity}
The complexity of AD with the modified MMV-AMP algorithm with ML estimation of LSFCs is in the order of $\mathcal{O}(MN\log N)$
when the pilots are chosen as the columns of a randomly sub-sampled DFT matrix (\cite{Fen2021e}) and
the approximate calculation of the
derivatives in the MMV-AMP as described in \cite{Fen2021a}.
For MMV-AMP with approximated PME the complexity is multiplied by the number of levels used to approximate the
distribution of LSFCs.
The complexity of MRC is $\mathcal{O}(K_aMn_d)$ and the complexity of single-user
decoding is $\mathcal{O}(K_an_d\log n_d)$ regardless of whether SIC has been used or not.
The SIC only influences the ability to process the users in parallel.

\section{Conclusion}
In this work we presented a coding scheme for the quasi-static Rayleigh fading AWGN
MAC with a massive MIMO receiver that is compliant with the unsourced paradigm.
I.e. no
user identification is done at the physical layer, instead the receiver recovers a list
of transmitted messages up to permutation.
Simulations show that the presented scheme
performs better than existing approaches for the quasi-static MIMO MAC despite its conceptual
simplicity. We give a closed form analysis that predicts the achievable limits of this scheme when
a single user code is used which can achieve the normal approximation.
We introduced a modified version of the MMV-AMP algorithm which allows to treat the unknown LSFCs in
a non-Bayesian way while maintaining the same complexity and still achieving results close to the Bayesian version without the need to obtain
the distribution of LSFCs. 
The results show that the presented approach can scale to many hundreds of users and achieve
high sum-spectral efficiencies with low power requirement and without any
coordination between users. Simplified power control policies allow to improve the performance even further
with only few interference cancellation steps being required at the receiver.
Specifically, with $M=50$ receive antennas and 3-stage SIC more than a thousand users
can be served concurrently,
which leads to sum-spectral efficiencies beyond 30 bits per channel use.

\printbibliography

\end{document}